\documentclass[prl, reprint, superscriptaddress, amsmath, amssymb, aps]{revtex4-2}

\usepackage{graphicx}
\usepackage{dcolumn}
\usepackage{bm}
\usepackage[protrusion=true,expansion=true]{microtype}
\usepackage[normalem]{ulem}
\usepackage{soul}
\usepackage{xcolor}
\usepackage{kotex}
\usepackage[shortcuts]{extdash}

\usepackage{cleveref}
\usepackage{bibunits}
\usepackage{float}

\defaultbibliographystyle{apsrev4-2}
\defaultbibliography{ref}

\usepackage{comment}

\newcommand{\adhoc}{{\it{ad hoc }}}
\def\Adhoc{{\bf \adhoc }}

\begin{document}

\begin{bibunit}

\preprint{APS/123-QED}

\title{
Explosive Synchronization in Inertial Kuramoto Models: \\ Determining Resilience in Power Grids}

\author{Cook Hyun Kim}
\affiliation{CCSS, KI for Grid Modernization, Korea Institute of Energy Technology, Naju, Jeonnam-Gwangju 58330, Korea}

\author{Jihye Kim}
\affiliation{CCSS, KI for Grid Modernization, Korea Institute of Energy Technology, Naju, Jeonnam-Gwangju 58330, Korea}

\author{Sangjoon Park}
\affiliation{CCSS, KI for Grid Modernization, Korea Institute of Energy Technology, Naju, Jeonnam-Gwangju 58330, Korea}


\author{J\"urgen Kurths}
\affiliation{Potsdam Institute for Climate Impact Research, Telegraphenberg, D-14415, Potsdam, Germany}

\author{B. Kahng}
\email{bkahng@kentech.ac.kr}
\affiliation{CCSS, KI for Grid Modernization, Korea Institute of Energy Technology, Naju, Jeonnam-Gwangju 58330, Korea}

\date{\today}

\begin{abstract}
Maintaining power\-/grid stability is essential for a sustainable energy supply. Using empirical power\-/grid data, we show that the inertial Kuramoto model with correlated powers and coupling strengths provides a relevant framework for analyzing explosive synchronization and grid stability. We identify two distinct synchronization pathways, initiated by population\-/driven or coupling\-/driven seed rotors, that yield different order\-/parameter behaviors. Inertia, in turn, determines whether the resulting seed clusters retain distinct angular velocities. Similar mechanisms govern recovery from external disturbances, linking power\-/grid resilience to these two nucleation pathways. Our findings provide a microscopic framework for understanding synchronization and recovery, with implications for stable power\-/grid operation.
\end{abstract}

\maketitle

\emph{Introduction}: A sustainable energy supply depends on stable electrical power grids. For a power grid to remain stable, all synchronous generators must share a common angular frequency. Transmission\-/line faults, generator tripping, and fluctuations in renewable output can separate these frequencies, and their recovery requires restoring the common frequency. Grid stability can therefore be viewed as the stability of a synchronized dynamical state, and recovery as the re\-/establishment of synchronization. Rotor dynamics follow the swing equation~\cite{kundur2007power,machowski2020power,witthaut2022collective,lee2024reinforcement,park2025optimal}, which is mathematically equivalent to the second\-/order Kuramoto model (2nd KM)~\cite{tanaka1997first,tanaka1997self,olmi2014hysteretic,gao2018self,gao2021synchronized,kim2025cluster,kim2026paths}. Understanding synchronization in the 2nd KM is therefore central to understanding power\-/grid stability.

The swing equation is derived from the balance of power in electric grids. In power grids, buses with larger power magnitudes tend to have greater total coupling strengths through their incident transmission lines. This correlation motivates the use of the explosive Kuramoto models~\cite{gomez2011explosive}, in which the natural frequencies are  correlated with the node degrees~\cite{peron2012determination,coutinho2013kuramoto,zou2014basin} or coupling strengths~\cite{zhang2013general}. Related studies consider frequency\-/dependent weighted coupling~\cite{leyva2013weighted} and extend frequency\=/degree or frequency\=/coupling correlations to the 2nd KM~\cite{ji2013cluster,ji2014analysis,peron2015explosive}. Here, we study such second\-/order explosive Kuramoto models (2nd EKM) to explore how synchronized clusters nucleate and propagate, and how these mechanisms relate to power\-/grid stability.

Ji et al.~\cite{ji2013cluster,ji2014analysis} studied a 2nd EKM on a heterogeneous network with correlated natural frequencies and node degrees, focusing mainly on the macroscopic synchronization transition. Here, we take a substantial step forward and use an all\-/to\-/all representation with heterogeneous coupling weights to resolve the microscopic nucleation, propagation, and recovery of synchronized clusters.

Inertia gives the 2nd KM microscopic dynamics that differ substantially from those of the first\-/order Kuramoto model (1st KM). Multiple synchronized clusters can nucleate with distinct angular velocities. Their coexistence and interactions produce a complex dynamical organization, including a Devil's staircase~\cite{kim2025cluster,kim2026paths}. Transient propagation, therefore, shapes the final synchronized state. These inertial dynamics call for a paradigm shift toward understanding synchronization transitions and cluster stability through the dynamical pathways of cluster formation. We consider power\-/law forms for the power distribution and coupling\-/weight profile and investigate how their correlation controls synchronization.

\begin{figure}[!t]
\centering
\includegraphics[width=1.0\linewidth]{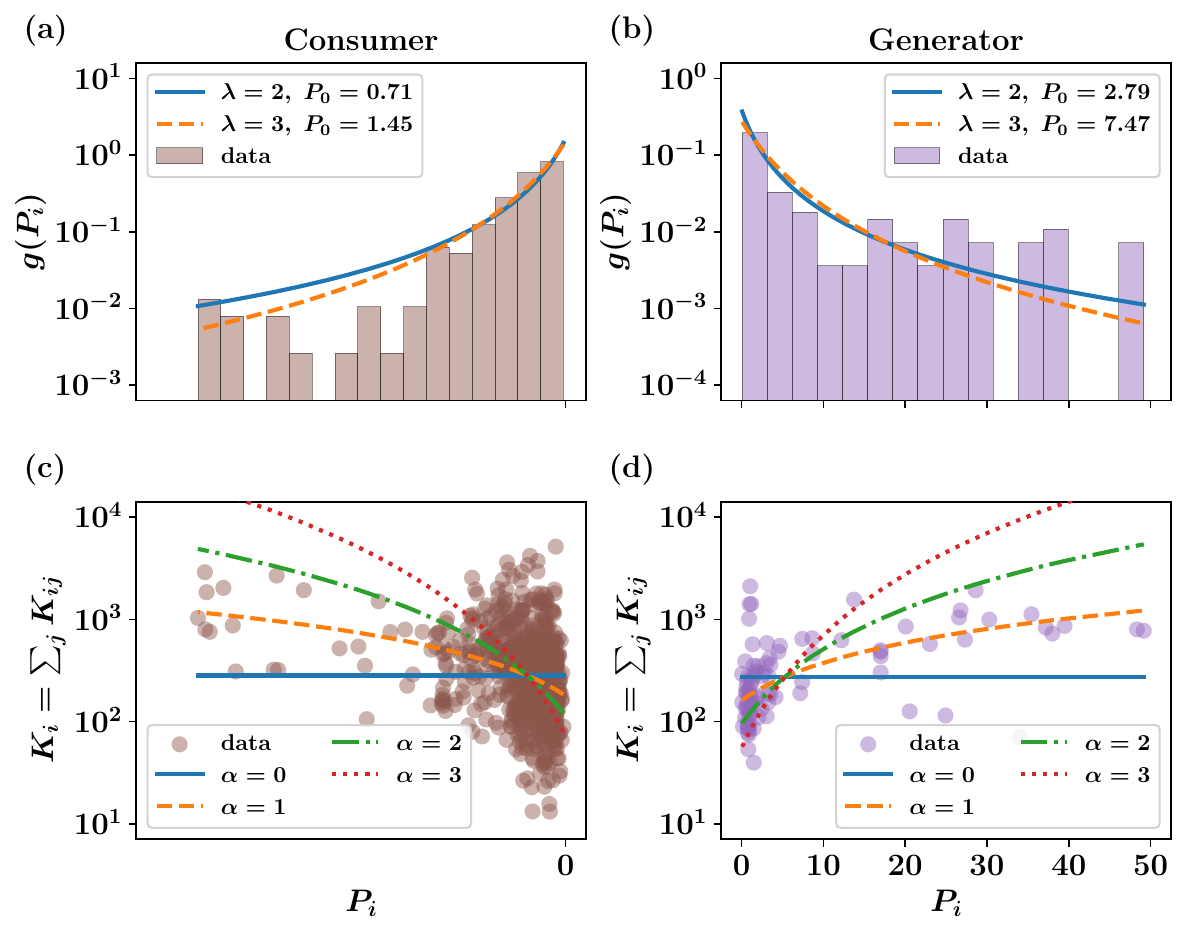}
\caption{\textbf{Power distribution and total coupling strength of the French transmission grid.} Node data from the French subgraph of the PanTaGruEl model~\cite{pagnier2019inertia}, with consumer buses (a,c) and generator buses (b,d) treated separately. (a,b)~Power distribution $g(P)$ (bars), with $(|P_i|+P_0)^{-\lambda}$ for $\lambda=2$ (solid) and $\lambda=3$ (dashed). The fitted $P_0$ for each curve is given in the legend. Each class spans more than one order of magnitude in $|P_i|$. (c,d)~Total coupling strength $K_i=\sum_j K_{ij}$ versus power (points), with $(|P_i|+P_0)^{\alpha}$ for $\alpha=0$--$3$ (lines), normalized by the mean. An exponent of order unity captures the observed trend; the uncorrelated case $\alpha=0$, for which $K_i$ is independent of $P_i$, does not.
}
\label{fig:fr_data}
\end{figure}

\begin{figure}[!t]
\centering
\includegraphics[width=1.0\linewidth]{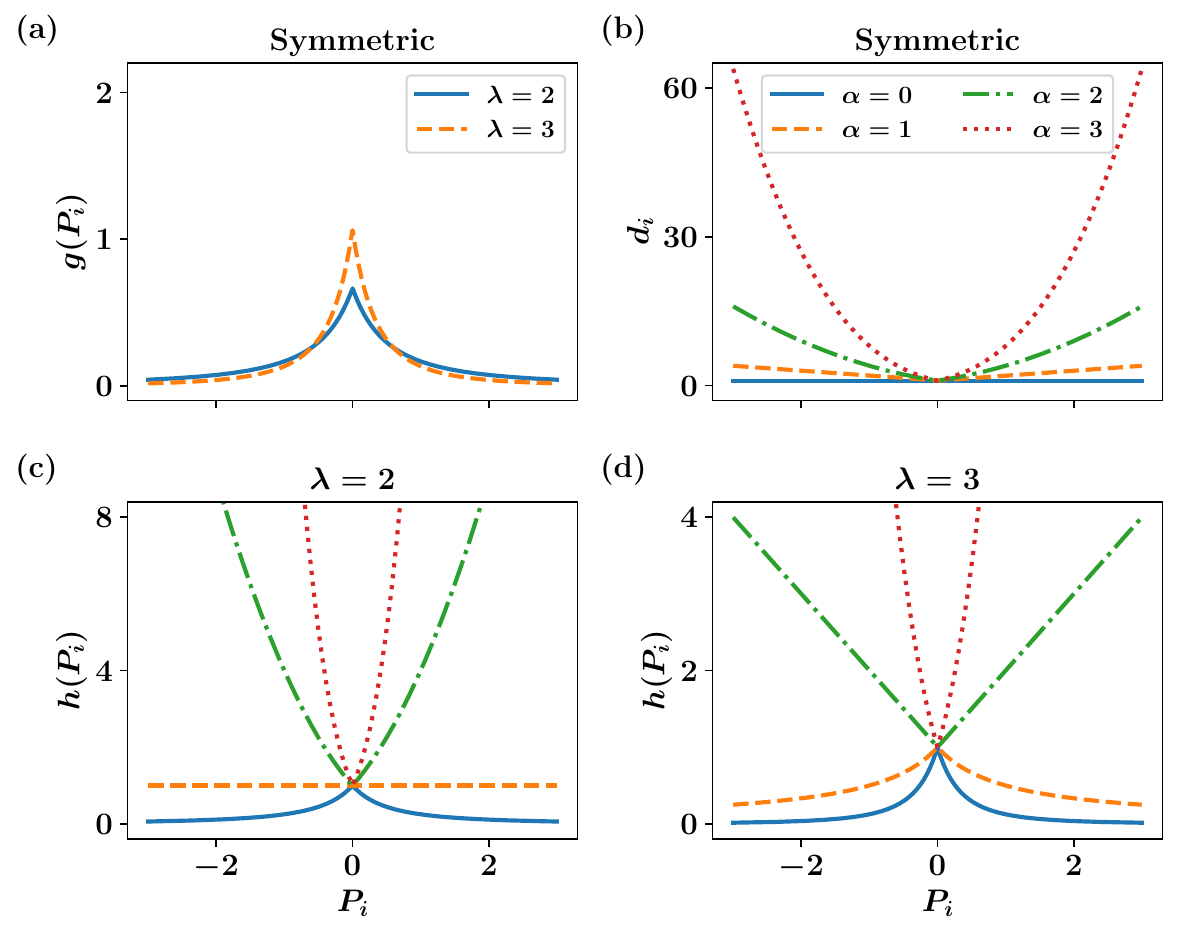}
\caption{\textbf{Rotor population, coupling\-/weight profile, and nucleation strength versus power $P$.}
(a)~Power distribution $g(P)$ for $\lambda=2,3$, peaked at $P=0$, with $P_0=0$ for illustration.
(b)~Coupling\-/weight profile $d(P)$ for $\alpha=0,1,2,3$, increasing toward both tails.
(c,d)~Nucleation strength $h(P)$ for $\lambda=2$ (c) and $\lambda=3$ (d). For $\alpha<\lambda/2$, $h(P)$ peaks at $P=0$ (population\-/driven); for $\alpha>\lambda/2$, $h(P)$ increases toward the tails (coupling\-/driven). The two contributions balance at $\alpha_c=\lambda/2$. For $\lambda=3$, $\alpha=1$ and $\alpha=2$ lie equally far from $\alpha_c=1.5$ on opposite sides.}
\label{fig:fig0}
\end{figure}

\begin{figure*}[!t]
\centering
\includegraphics[width=1.0\linewidth]{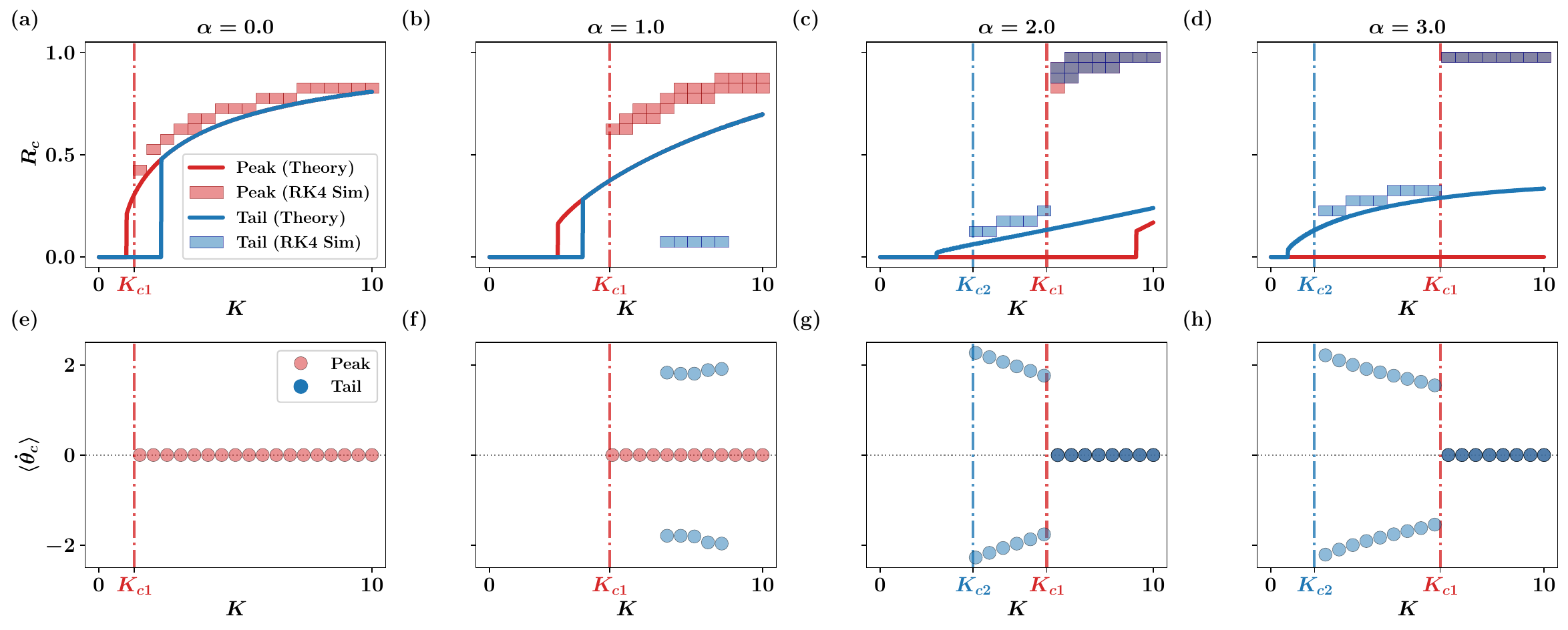}
\caption{\textbf{Population\-/driven and coupling\-/driven synchronization (symmetric distribution, $m=10$).}
(a--d)~$R_c$ versus $K$ for $\alpha=0,1,2,3$. Solid curves: population\-/driven (red) and coupling\-/driven (blue) pathways predicted by the \Adhoc potential. Squares: the most populated $R_c$ bins over $64$ realizations at each $K$ for each pathway. Dash\-/dotted vertical lines: predicted onsets $K_{c1}$ (red) and $K_{c2}$ (blue).
(e--h)~Modal positions of the mean cluster angular velocities $\langle\dot\theta_c\rangle$ for the same $\alpha$. For $\alpha<\alpha_c$, a central\-/seed cluster nucleates near $\langle\dot\theta_c\rangle=0$ and grows by accretion. For $\alpha>\alpha_c$, two tail\-/seed clusters nucleate at finite $\pm\langle\dot\theta_c\rangle$ and merge into a central cluster.
The \Adhoc potential reproduces $R_c(K)$ and selects the pathway observed in the simulations. The corresponding potential profiles are shown in Fig.~\ref{fig:sym_potential} in the SM.}
\label{fig:sym_massive_Rc}
\end{figure*}

\emph{Model.}---The dynamics of a power grid are described by the swing equation,
\begin{align}
m\ddot{\theta}_i+\gamma\dot{\theta}_i
=
P_i+\sum_j K_{ij}\sin(\theta_j-\theta_i),
\end{align}
where $m$ is the inertia, $\gamma$ is the damping coefficient, $\theta_i$ is the phase at bus $i$, $P_i$ is its power, and $K_{ij}$ is the line coupling strength between buses $i$ and $j$. Dots denote time derivatives.

Inertia and damping generally vary across buses and may depend on $P_i$. We take $m$ and $\gamma$ as constants to isolate the effects of the power distribution and the power\=/coupling correlation. In real power grids, the total coupling strength of bus $i$,
\begin{align}
K_i\equiv\sum_{j\in\mathrm{neighbors}(i)}K_{ij},
\end{align}
can be correlated with its power $P_i$, as observed in the French grid~\cite{pagnier2019inertia}. We incorporate this correlation into the 2nd EKM,
\begin{align}
m\ddot{\theta}_i + \gamma\dot{\theta}_i
&=
P_i
+
\frac{K d_i}{\langle d\rangle}
\frac{1}{N}
\sum_{j=1}^{N}
\frac{d_j}{\langle d\rangle}
\sin(\theta_j-\theta_i),
\label{eq:model}
\end{align}
where $K$ is the global coupling strength and $N$ is the number of rotors. The heterogeneous total coupling strength is
\begin{align}
K_i=K\frac{d_i}{\langle d\rangle},
\end{align}
where $d_i$ is the relative coupling weight of rotor $i$.

The powers $\{P_i\}$ are drawn from $g(P)\propto(|P|+P_0)^{-\lambda}$ on $[-P_b,P_b]$, and the coupling\-/weight profile is $d(P)=(|P|+P_0)^{\alpha}$, with $\langle d\rangle=N^{-1}\sum_i d_i$. The power $P_i$ plays the role of the natural frequency in the standard Kuramoto model. The regularizer $P_0$ removes the singularity of $g(P)$ at $P=0$. Simulations use $P_0=1$; the scaling arguments below neglect $P_0$ when $P_0\ll|P|$.

The forms of $g(P)$ and $d(P)$ are motivated by the French subgraph of the PanTaGruEl model of the European transmission system~\cite{pagnier2019inertia}, with consumer and generator buses treated separately. Each class spans more than one order of magnitude in $|P_i|$. The exponent $\lambda$ is approximately $2$--$3$ [Figs.~\ref{fig:fr_data}(a,b)], and $\alpha$ is of the order of unity [Figs.~\ref{fig:fr_data}(c,d)]. Larger $|P_i|$ therefore tends to be associated with a larger total coupling strength $K_i$. Equation~\eqref{eq:model} provides a minimal model of this correlation. The relative magnitudes of the exponents determine the propagation pathway; the precise exponent values are not required.

For $\alpha=0$, the coupling weights are uniform and Eq.~\eqref{eq:model} reduces to the standard uncorrelated 2nd KM. Increasing $\alpha$ gives a greater relative coupling weight to high-$|P|$ rotors. The exponents $\lambda$ and $\alpha$ therefore favor opposite regions: $\lambda$ concentrates the rotor population near the center of the power distribution [Fig.~\ref{fig:fig0}(a)], while $\alpha$ concentrates relative coupling weight toward the tails [Fig.~\ref{fig:fig0}(b)]. Their competition determines where synchronization nucleates.

A synchronized seed cluster can nucleate among rotors with similar powers. A power bin contains a rotor population proportional to $g(P)$, and each rotor has the relative coupling weight $d(P)$. The bin therefore generates a local mean field with a total weight proportional to $g(P)d(P)$, while each rotor responds with the relative coupling weight $d(P)$. We define the product as the nucleation strength,
\begin{align}
h(P)\;\equiv\; d(P)\!\sum_{j\in\text{bin}} d_j
\;\sim\; g(P)\,d(P)^{2}
\;\sim\;(|P|+P_0)^{\,2\alpha-\lambda}.
\label{eq:heff}
\end{align}
The sign of $2\alpha-\lambda$ determines where $h(P)$ is maximal [Figs.~\ref{fig:fig0}(c,d)]. i) For $\alpha<\lambda/2$, $h(P)$ is maximal at $P=0$: a seed cluster nucleates at the high\-/population center, and synchronization propagates toward the tails. We call this the {\it population\-/driven pathway}. ii) For $\alpha>\lambda/2$, $h(P)$ is maximal at the tails: seed clusters nucleate among strongly coupled high-$|P|$ rotors, and synchronization propagates toward the center. We call this the {\it coupling\-/driven pathway}. These two regimes meet at the balance point $\alpha_c=\lambda/2$. We fix $\lambda=3$, giving $\alpha_c=1.5$, and study $\alpha\in\{0,1,2,3\}$.

We quantify the coherence of synchronized cluster $c$ by the weighted cluster order parameter $R_c$, defined by $R_c e^{\mathrm{i}\Theta_c}=\sum_{i\in c}(d_i/\sum_{j\in c}d_j)e^{\mathrm{i}\theta_i}$. Here, $\Theta_c$ is the mean cluster phase, and the sums run over the cluster members.

\begin{figure*}[!t]
\centering
\includegraphics[width=1.0\linewidth]{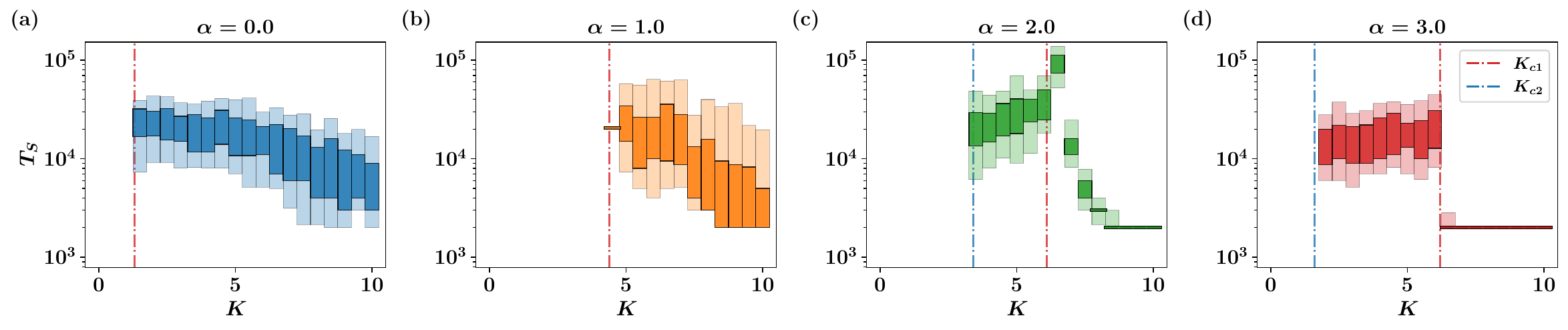}
\caption{\textbf{Transient\-/time distributions distinguish accretion from merger (symmetric distribution, $m=10$).} (a--d)~$T_S$ over $64$ realizations versus $K$ for $\alpha=0,1,2,3$. Light and dark boxes span $Q_{5}$--$Q_{95}$ and $Q_{25}$--$Q_{75}$, respectively; when $Q_{25}=Q_{75}$, a horizontal bar marks the median (SM). All panels share a logarithmic vertical axis from the fine\-/window resolution $10^{3}$ to $1.2\times10^{5}$ time units. Quantiles below $10^{3}$ are unresolved and omitted; boxes extending above $1.2\times10^{5}$ are clipped. Panels (a--c) cover $K\le10$, and panel (d) covers $K\le20$. Dash\-/dotted vertical lines: measured onsets $K_{c1}$ (red) and $K_{c2}$ (blue). Population\-/driven regime ($\alpha=0,1$): $T_S$ remains large and broadly distributed above $K_{c1}$ as the macroscopic cluster entrains high-$|P|$ rotors one at a time. Coupling\-/driven regime ($\alpha=2,3$): $T_S$ is broadly distributed for $K_{c2}<K<K_{c1}$, where tail\-/seed clusters coexist before merger, and becomes small above $K_{c1}$. In this regime, $K_{c1}$ marks the emergence of the merged macroscopic cluster.}
\label{fig:critical}
\end{figure*}

\emph{Nucleation bin and persistence.}---The cluster\-/resolved analysis identifies the nucleation bin [Fig.~\ref{fig:sym_massive_Rc}]. In the population\-/driven regime (e.g.,~$\alpha=0,1$), a central\-/seed cluster nucleates near $\langle\dot\theta_c\rangle=0$ at $K_{c1}$ and grows by accretion, entraining rotors one at a time in decreasing order of $h(P)$. In the coupling\-/driven regime (e.g.,~$\alpha=2,3$), two tail\-/seed clusters nucleate at finite $\pm\langle\dot\theta_c\rangle$ at $K_{c2}$. The tails contain few rotors, so the initial increase in $R_c$ is small. Additional mesoscopic clusters nucleate and merge with the tail\-/seed clusters, producing stepwise increases in $R_c$. The two macroscopic clusters finally merge at $K_{c1}$.

To describe the propagation pathways, we introduce the \Adhoc potential $\mathcal{A}(R_c)$~\cite{song2020effective,jhun2022quantum}, derived from the self\-/consistency equation for the cluster order parameter $R_c$ [Supplemental Material (SM)~\cite{SM}]. The potential is accumulated along a propagation path as the cluster center $P_c$ follows the mean power of the cluster members. The population\-/driven path starts at the center of the power distribution, and the coupling\-/driven path starts at a tail. At a given $K$, a minimum of $\mathcal{A}(R_c)$ identifies a stable cluster, and the lower of the two pathway minima identifies the selected pathway [Fig.~\ref{fig:sym_potential} in the SM]. The gap $\Delta\mathcal{A}$ measures, how strongly the selected pathway is favored. Allowing $P_c$ to evolve makes the construction applicable to both symmetric and asymmetric power distributions.

The \Adhoc potential favors the population\-/driven pathway for $\alpha=0,1$ and the coupling\-/driven pathway for $\alpha=2,3$, consistent with the sign of $2\alpha-\lambda$ and the simulations [Figs.~\ref{fig:sym_massive_Rc}(a--d)]. In the coupling\-/driven regime, the potential also captures the tail\-/seed onset at $K_{c2}$ and the small initial $R_c$.

The maximum of the nucleation strength $h(P)$, determined by the $g(P)$ and $d(P)$, selects the nucleation bin at the center or tails of the power distribution. Inertia controls whether distinct seed clusters persist. Together, selection and persistence distinguish single\-/stage population\-/driven propagation from a multistage coupling\-/driven propagation.

The same selection rule holds in the first\-/order EKM when $\alpha$, $g(P)$, and $d_i$ are unchanged. The nucleation bins remain the same, but the seed clusters merge rapidly into a macroscopic cluster near the center of $g(P)$ (SM). Inertia therefore changes persistence and propagation without changing the nucleation bin.

We next measure the transient time $T_S$ required for the cluster structure to reach its final form over $64$ realizations at each $K$ (SM). In the population\-/driven regime, $T_S$ decreases as $K$ increases above $K_{c1}$ [Fig.~\ref{fig:critical}(a,b)]. In the coupling\-/driven regime ($\alpha=2,3$), $T_S$ increases with $K$ up to $K_{c1}$. The peak is higher for $\alpha=2$ than for $\alpha=3$ because $\alpha=2$ lies closer to the balance point $\alpha_c=1.5$. Near $\alpha_c$, the two pathways become comparable, allowing nucleation near both the center and tails of the power distribution. The central\-/seed cluster is attracted by both tail\-/seed clusters and moves back and forth between them. This motion prolongs the transient, analogous to slowing down near a saddle\-/node bifurcation. For $K>K_{c1}$, $T_S$ decreases rapidly as merger produces a macroscopic cluster [Fig.~\ref{fig:critical}(c,d)].

\emph{Robustness and resilience of the synchronized state.}---Linear stability alone does not capture the stability of a synchronized state against finite perturbations~\cite{menck2013basin,Menck2014}. We perturb a synchronized state by selecting $256$ of the $1024$ rotors with the largest relative coupling weights $d_i$ (equivalently, the largest $|P_i|$ in the idealized model) and resetting their angular velocities to the free\-/running values $P_i/\gamma$. We measure the largest cluster size $S_{\max}$ to quantify robustness and the recovery time $T_R$ to quantify resilience (SM) and fix $m=10$.

Recovery follows the same propagation pathway because the perturbed rotors retain their coupling weights. i) In the population\-/driven regime, the central core survives and entrains the detached tail rotors one at a time, so recovery proceeds by accretion. ii) In the coupling\-/driven regime, the perturbed tail rotors are the original strongly coupled seed rotors. These rotors nucleate a new cluster that joins the surviving core, so recovery proceeds by merging.

\begin{figure*}[!t]
\centering
\includegraphics[width=1.0\linewidth]{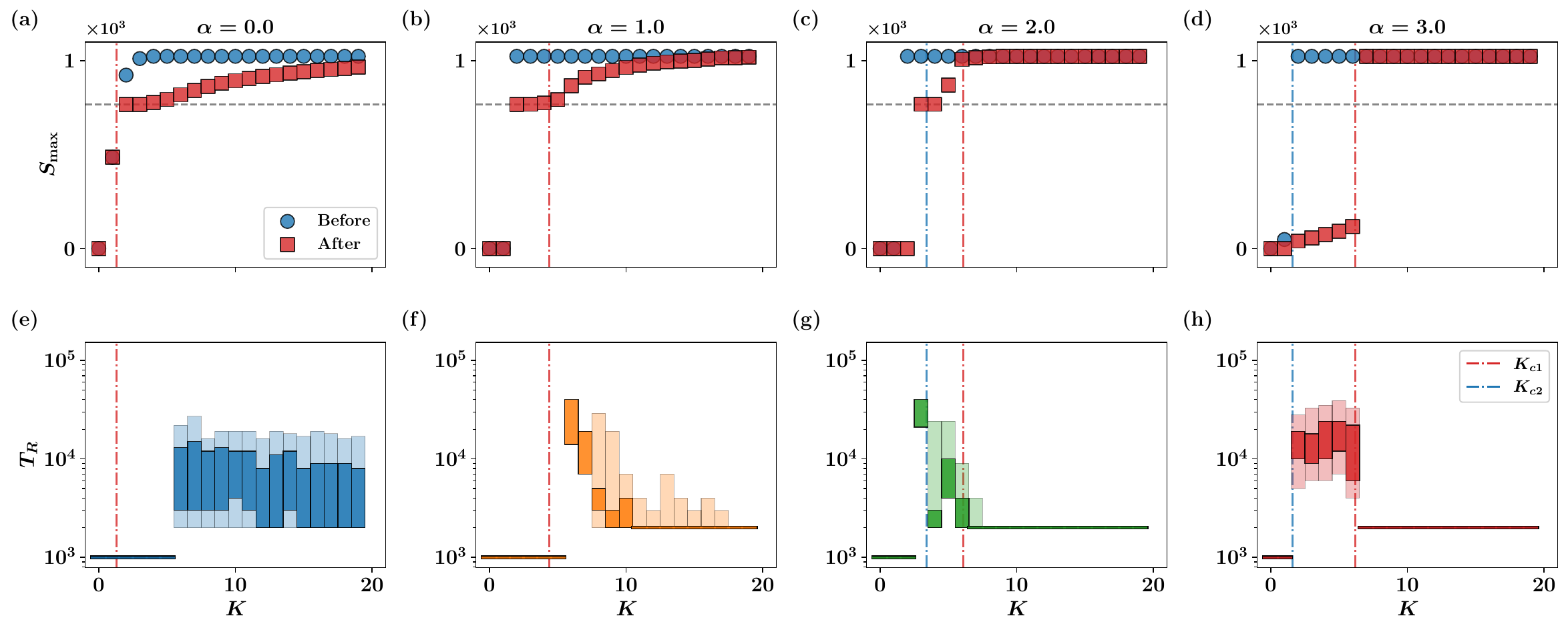}
\caption{\textbf{Robustness and resilience of the synchronized state (symmetric distribution, $m=10$).} (a--d)~Largest cluster size $S_{\max}$ before (blue) and after (red) the perturbation versus $K$, reported as medians over $64$ realizations (SM). Population\-/driven regime ($\alpha=0,1$): a macroscopic core survives at every $K$ above onset, and recovery proceeds gradually by accretion. Coupling\-/driven regime ($\alpha=2,3$): no macroscopic cluster survives at low $K$, while merger fully restores the unperturbed cluster size at high $K$. (e--h)~Recovery time $T_R$ over the same realizations, with the box construction used in Fig.~\ref{fig:critical}. The quantiles include all $64$ realizations. A quantile beyond the maximum recovery time $5\times10^{4}$ is censored, and the corresponding box is omitted. Dash\-/dotted vertical lines: measured onsets $K_{c1}$ (red) and $K_{c2}$ (blue).}
\label{fig:stability}
\end{figure*}

These two pathways produce distinct $S_{\max}$ and $T_R$ [Fig.~\ref{fig:stability}]: Population\-/driven recovery preserves a macroscopic core throughout the synchronized range, but restores detached rotors gradually. Coupling\-/driven recovery fails to preserve a macroscopic cluster at weak coupling, but fully restores the unperturbed cluster once coupling is strong enough for merger.

\emph{Generality beyond the idealized distributions.}---We find that the same selection\=/persistence mechanism even appears for more complex power distributions and coupling\-/weight profiles. We also demonstrate that data\-/derived annealed representations of the French, German, Spanish, and British grids reproduce these two key features: the maxima of $h(P)$ identify the nucleation bins, and inertia controls whether the resulting seed clusters persist with distinct angular velocities. These results clearly extend the mechanism beyond the idealized power\-/law distributions (SM).

\emph{Discussion.}---
We have investigated synchronization in the 2nd EKM by resolving the nucleation and propagation of synchronized clusters. The nucleation strength $h(P)\sim g(P)d(P)^2$ identifies the nucleation bin and propagation pathway: population\-/driven propagation from the center of the power distribution or coupling\-/driven propagation from its tails. In the population\-/driven pathway, a central\-/seed cluster grows by rotor accretion, and $T_S$ decreases with increasing $K$ above $K_{c1}$. In the coupling\-/driven pathway for a symmetric distribution, tail\-/seed clusters grow by merger and form a single macroscopic cluster at $K_{c1}$.

Near the balance point $\alpha_c=\lambda/2$, the competition between these two pathways lengthens the transient time and broadens its distribution. The inertia controls the persistence of the selected seed clusters as follows: with inertia, the clusters retain distinct angular velocities over a broad range of $K$; without inertia, the clusters merge soon after nucleation and multistage propagation disappears.

These propagation mechanisms also govern the recovery from external perturbations. Analyses of asymmetric power distributions and data\-/derived annealed representations of European grids extend the selection\=/persistence mechanism beyond the symmetric idealized model (SM).

These findings extend the understanding of first\-/order Kuramoto models with frequency\=/coupling correlations. Previous work classified collective states, including traveling\-/wave and standing\-/wave states, and analyzed their onset and stability~\cite{iatsenko2013stationary,xu2016synchronization,bi2017nontrivial}. Our transient analysis interprets these states as coupling\-/driven seed\-/cluster configurations, whose persistence is controlled by inertia: an asymmetric distribution produces one rotating seed cluster, while a symmetric distribution produces a counter\-/rotating pair. The \Adhoc potential selects the propagation pathway and predicts the coupling at which its seed cluster nucleates.

Allowing the cluster center $P_c$ to evolve, the \Adhoc potential construction captures nucleation at the tail of $g(P)$ as well as at its center. The pathway depth $\mathcal{A}(R_c^*)$ ranks competing seed clusters, and the gap $\Delta\mathcal{A}$ also measures the validity of the single\-/cluster description. As $\alpha\to\alpha_c$, the two pathways approach equal depth, inter\-/cluster interactions become important, and the coexistence interval broadens. This construction can also be applied when the power or natural\-/frequency distribution, coupling weights, and entrainment radius are specified. Examples include systems with heterogeneous or adaptive inertia and damping~\cite{choi2019synchronization,nnoli2021spreading,fritzsch2024stabilizing,park2025optimal,kim2026decentralization}; systems with network\-/structured or multilayer coupling~\cite{oh2005modular,park2006synchronization,gomez2007paths,zhang2015adaptive,kachhvah2017multiplexing,odor2018heterogeneity}; and systems with frequency\-/dependent coupling~\cite{zhang2013general,leyva2013weighted,xu2016synchronization,qiu2016synchronization}, time\-/delayed feedback~\cite{taher2019enhancing}, or phase lag~\cite{yi2026phase}.

Ji et al.~\cite{ji2013cluster} correlated the natural frequency with the node degree in a second\-/order Kuramoto model. We correlate power with total coupling strength in an all\-/to\-/all system. Their linear frequency\=/degree correlation makes synchronization propagate from the high\-/population, low\-/degree region, corresponding to our population\-/driven pathway. A sufficiently superlinear correlation could shift the nucleation toward the low\-/population, high\-/degree region, corresponding to the coupling\-/driven pathway. Their cluster explosive synchronization can therefore be interpreted as a particular case of the propagation mechanisms identified here.

The results are also relevant to grids with high renewable penetration. Replacing synchronous generation with inverter\-/based resources reduces rotational inertia and shifts the grid toward a low\-/inertia operation~\cite{milano2018foundations,tayyebi2020frequency,Sajadi2022}. In our framework, replacing a few large synchronous plants with many smaller generating units also shifts weight away from the high\-/power tail of the power distribution. This shift increases the population contribution relative to the coupling contribution in $h(P)$. These two changes affect different stages: the shift in the power distribution favors a central, population\-/driven seed, while the reduction in inertia suppresses the persistence of distinct tail\-/seed clusters. Both changes therefore favor a single synchronized cluster nucleated near the center of the power distribution. The nucleation strength and inertia thus organize selection and persistence in both synchronization and recovery.

More broadly, our pathway\-/based framework resolves how clusters nucleate, persist, propagate, and recover, linking formation dynamics to synchronization transitions and cluster stability. These findings call for a paradigm shift toward a dynamical understanding of synchronization centered on propagation pathways, extending static\-/state criteria to the processes that establish and restore synchronization.

\begin{acknowledgments}
B.K. was supported by the National Research Foundation of Korea (Grant No. RS-2023-00279802 and No. RS-2026-25556154), and the KENTECH Research Grant No. KRG-2021-01-007.  
\end{acknowledgments}







\putbib
\end{bibunit}

\clearpage
\newpage

\appendix

\onecolumngrid

\makeatletter
\renewcommand{\thesection}{S\arabic{section}}
\renewcommand{\theequation}{S\arabic{equation}}
\renewcommand{\thefigure}{S\arabic{figure}}
\renewcommand{\thetable}{S\arabic{table}}
\renewcommand{\bibnumfmt}[1]{[S#1]}

\setcounter{section}{0}
\setcounter{equation}{0}
\setcounter{figure}{0}
\setcounter{table}{0}
\setcounter{page}{1}
\makeatother

\begin{center}
\textbf{\large Supplemental Material for\\
``Explosive Synchronization in Inertial Kuramoto Models: \\ Determining Resilience in Power Grids''}
\end{center}

\begin{bibunit}

\section{Numerical methods}
\label{sm:numerics}

All simulations use a fourth\-/order Runge\=/Kutta (RK4) scheme. Unless stated otherwise, observables are time\-/averaged in the steady state, and ensemble statistics are calculated over $64$ independent realizations.

\subsection{Idealized model}
\label{sm:model_numerics}

We integrate Eq.~\eqref{eq:model} for $N=1024$ rotors in the all\-/to\-/all representation with $\Delta t=10^{-2}$ and $\gamma=1$. Runs with inertia use $m=10$. Runs without inertia use $m=0$, reducing the dynamics to $\gamma\dot\theta_i=P_i+(Kd_i/\langle d\rangle)N^{-1}\sum_j\sin(\theta_j-\theta_i)$.

We sample $K$ over $[0,20)$ in steps of $0.1$. Each $K$ is simulated independently, with phases initialized uniformly on $[0,2\pi)$ and, for runs with inertia, angular velocities initialized at $\dot\theta_i=P_i/\gamma$. No state is carried between neighboring values of $K$. Variation at fixed $K$ therefore reflects sensitivity to initial conditions, with no hysteresis introduced by continuation in $K$.

Powers and relative coupling weights are generated by deterministic sampling followed by a random permutation. A given seed therefore reproduces the same $\{P_i\}$ in simulation and post\-/processing. For the symmetric distribution, powers are sampled from $g(P)\propto(|P|+P_0)^{-\lambda}$ on $[-P_b,P_b]$ with $P_0=1$ and $P_b=4$, and the weights $d_i=(|P_i|+P_0)^{\alpha}$ are normalized by $\langle d\rangle$. For the asymmetric distribution, raw powers $P_i^{\rm raw}$ are sampled from the same power law on $[0,P_b]$ and mean\-/centered to give $P_i=P_i^{\rm raw}-\langle P^{\rm raw}\rangle$; weights $d_i$ are computed from the pre\-/shift powers $P_i^{\rm raw}$. The smallest and largest sampled powers are denoted by $P_{\rm lo}$ and $P_{\rm hi}$. We fix $\lambda=3$, giving $\alpha_c=1.5$, and study $\alpha\in\{0,1,2,3\}$.

The measurement protocol has two stages. First, we advance the dynamics in segments until the cluster structure becomes stationary (Sec.~\ref{sm:clustering}), defining the transient time $T_S$ (Sec.~\ref{sm:tc}). Second, we advance the system for another $10^3$ time units and time\-/average $R_c$ and $\langle\dot\theta_c\rangle$ over this interval.

\subsection{European grids}
\label{sm:grid_numerics}

\paragraph{Data and national subgraphs.}
The grid data come from the PanTaGruEl model of the synchronous transmission grid of continental Europe~\cite{pagnier2019inertia}, available under a CC\-/BY-4.0 license~\cite{pantagruel_data}. Each bus has a raw power $P_i^{\rm raw}$ (generation minus load), and each transmission line has a coupling strength. We extract four national subgraphs (DE, ES, FR, UK), retaining each country's buses and internal lines and removing cross\-/border ties. We then restore power balance so that $\sum_i P_i^{\rm raw}=0$. The subgraphs retain the generator\-/heavy tails of the original networks.

\paragraph{Annealed reduction and rescaling.}
Each subgraph is mapped onto Eq.~\eqref{eq:model} in the annealed limit. Empirical powers are rescaled to the power range of the idealized model: generators are mapped to the positive\-/power branch, and consumers to the negative\-/power branch. The rescaled powers are then mean\-/centered to give $P_i$. This mapping preserves the shape of each branch and places the empirical and idealized systems on the same scale.

The relative coupling weight is constructed from the rescaled power $P_i$, the raw power $P_i^{\rm raw}$, and the total coupling strength $\kappa_i$ (the sum of incident line admittances) as
\begin{equation}
d_i=(1+|P_i|)^{\alpha}\,w_i,
\qquad
w_i=\frac{\kappa_i}{\sum_j \kappa_j}\sum_j\big(1+|P_j^{\rm raw}|\big)^{\alpha},
\end{equation}
with the same $\alpha$ in both factors and normalization $\langle d\rangle=1$. At $\alpha=0$, $d_i\propto\kappa_i$, so the relative coupling weights reproduce the total coupling strengths. Increasing $\alpha$ shifts relative coupling weight toward high-$|P|$ buses. The rescaling is independent of inertia; runs with and without inertia at a given $\alpha$ therefore differ only in $m$. We report $\alpha=3$ in Sec.~\ref{sm:grids} and use $\alpha=0$ as a consistency check.

\paragraph{Equations of motion.}
The rescaled powers $P_i$ and relative coupling weights $d_i$ define the annealed dynamics,
\begin{equation}
m_i\ddot\theta_i+\gamma\dot\theta_i
=P_i+K\,d_i\!\left(
\frac{\sum_j d_j\sin\theta_j}{\sum_j d_j}\cos\theta_i
-\frac{\sum_j d_j\cos\theta_j}{\sum_j d_j}\sin\theta_i\right),
\label{eq:grid_eom}
\end{equation}
which is the coupling\-/weighted mean\-/field form of Eq.~\eqref{eq:model}. All buses have $\gamma=1$; runs with and without inertia use $m_i=10$ and $m_i=0$, respectively. Phases are reinitialized independently at each $K$, and runs with inertia start from $\dot\theta_i=P_i/\gamma$. We sample $K\in[0,2]$ in steps of $0.01$. The time step is $\Delta t=10^{-2}$ with inertia and $\Delta t=2\times10^{-3}$ without inertia. A transient of $1500$ time units is followed by a measurement window of $500$ time units.

\subsection{Cluster detection}
\label{sm:clustering}

We identify clusters by one\-/dimensional clustering of the time\-/averaged angular velocities $\langle\dot\theta_i\rangle_t$. We first sort the values and group adjacent values whose separation is below an initial threshold. We then discard groups below the minimum size, iteratively merge neighboring groups within a merge tolerance, and assign remaining rotors to the nearest group within the same tolerance. Each detected cluster is characterized by its size $n_c$, mean angular velocity $\langle\dot\theta_c\rangle$, and mean power $P_c$.

For the idealized model, the initial squared\-/difference threshold is $10^{-8}$ (an angular\-/velocity gap of $10^{-4}$), the merge and assignment tolerance is $10^{-2}$, and the minimum cluster size is $32$. For the annealed grid representations, these values are $10^{-2}$, $1$, and $4$, respectively, reflecting their smaller sizes and broader angular\-/velocity ranges.

\subsection{Observables}
\label{sm:observables}

We report the cluster order parameter $R_c$ as a distribution. At each $K$, we divide $[0,1]$ into $20$ bins and show up to four bins whose occupancy exceeds $1/\sqrt{64}$. Two coexisting clusters are represented by two occupied $R_c$ bins.

We classify the pathways by cluster membership. For the symmetric distribution, a cluster is assigned to the population\-/driven pathway if it contains at least $16$ of the $32$ rotors with the smallest $|P_i|$, and to the coupling\-/driven pathway if it contains at least $16$ of the $32$ rotors with the largest $|P_i|$. A cluster may satisfy both criteria. For the asymmetric distribution, the corresponding reference sets contain the $32$ smallest and $32$ largest $P_i$. For the annealed grid representations, clusters are labeled by the sign of their mean power $P_c$. The angular velocities plotted in the figures are the modal positions of the $\langle\dot\theta_c\rangle$ values pooled across realizations.

\subsection{Transient time}
\label{sm:tc}

The transient time $T_S$ is the time required for the cluster structure to reach its final form. We average angular velocities over successive fine windows of $10^3$ time units and confirm stationarity using consecutive coarse windows of $10^4$ time units. A realization is classified as trivial if its largest cluster never reaches the minimum size.

At each $K$, we exclude trivial realizations and realizations that do not reach stationarity within $3.6\times10^{5}$ time units from the transient\-/time statistics. Figures~\ref{fig:critical} and~\ref{fig:asym_critical} show $Q_5$--$Q_{95}$ and $Q_{25}$--$Q_{75}$ as nested boxes.
 
The logarithmic vertical axis extends from the fine\-/window resolution $10^3$ to $1.2\times10^5$. Quantiles below the resolution are unresolved and omitted; boxes extending above the upper limit are clipped. When $Q_{25}=Q_{75}$, a horizontal bar marks the median.

\subsection{Data-derived $h(P)$ for the grid representations}
\label{sm:hP}

We evaluate the data\-/derived nucleation strength without fitting $\alpha$ or $\lambda$. We divide the power range into $16$ uniform bins and calculate $h(P)=n_{\rm bin}\langle d\rangle_{\rm bin}^2$ in each bin, where $n_{\rm bin}$ is the occupancy and $\langle d\rangle_{\rm bin}$ is the mean relative coupling weight. The power distribution is normalized as $\sum_{\rm bin}g_{\rm bin}\Delta P_{\rm bin}=1$. The maximum of $h(P)$ identifies the data\-/derived nucleation bin.

\clearpage
\newpage

\subsection{Perturbation protocol}
\label{sm:perturbation}

The robustness and resilience analysis starts from a synchronized operating state prepared with deterministic initial conditions. At each $K$, we independently prepare the operating state, perturb the tail rotors, and follow the relaxation.

\paragraph{Operating state.}
At each $K$, we integrate Eq.~\eqref{eq:model} for $10^{4}$ time units from the deterministic initial condition $\dot\theta_i=0$ and $\theta_i=\arctan(P_i/K)$ for $|P_i|\le K$, with $\theta_i=\pm\pi/2$ otherwise. This initial condition approximates a locked configuration. We then integrate for another $10^{3}$ time units to time\-/average the order parameters and detect clusters as in Sec.~\ref{sm:clustering}. The largest cluster defines the unperturbed $S_{\max}$ in Fig.~\ref{fig:stability}.

\paragraph{Perturbation.}
We perturb the $256=N/4$ rotors with the largest relative coupling weights $d_i$, corresponding to the high-$|P|$ region where coupling\-/driven seeds nucleate. Each selected rotor is assigned the free\-/running angular velocity $\dot\theta_i=P_i/\gamma$ and a phase drawn uniformly on $[0,2\pi)$; all other rotors retain their operating\-/state values. For the symmetric distribution, the perturbed set contains the largest-$|P_i|$ rotors from both tails. For the asymmetric distribution, the set contains the rotors with the largest positive $P_i$.

\paragraph{Relaxation.}
We integrate the perturbed state for at most $5\times10^{4}$ time units. The recovery time $T_R$ is the earliest fine window of width $10^3$ after which the cluster structure matches its final relaxed form, using the same stationarity criterion as for $T_S$. We then integrate for another $10^3$ time units to measure the relaxed order parameters and post\-/perturbation $S_{\max}$.

\paragraph{Ensemble.} We use $64$ realizations per $K$, with fixed $\{P_i\}$, $\{d_i\}$, and perturbed set, and sample $K=0,1,\dots,19$. We report the median $S_{\max}$. Recovery\-/time quantiles use the full ensemble, including censored realizations that do not relax within $5\times10^{4}$ time units. A quantile is omitted when too few realizations recover to determine its value.

\clearpage
\newpage

\section{\Adhoc potential}
\label{sm:adhoc}

An effective potential of a single macroscopic variable represents collective dynamics through an energy landscape. This approach is used across collective systems~\cite{jang2015ashkin,kim2024entropy,kim2026heterogeneous}. The \Adhoc potential method~\cite{song2020effective,jhun2022quantum} constructs the potential from the self\-/consistency equation for the order parameter. In the first\-/order KM, the potential describes the stability of a synchronized steady state. In the 2nd EKM, we use the potential to describe propagation\-/path selection during the transient.

\subsection{The propagation path and the \Adhoc potential}

Consider a synchronized cluster with order parameter $R_c$ and center $P_c$, defined as the mean power of its members. In the all\-/to\-/all representation, rotor $i$ has total coupling strength
\begin{align}
K_i=K\frac{d_i}{\langle d\rangle}.
\end{align}

With inertia, cluster membership is determined by the Melnikov entrainment radius,
\begin{align}
r_{c,i}
=
\frac{4\gamma}{\pi}
\sqrt{\frac{K_iR_c}{m}}
+
\frac{c_m}{\sqrt{K_iR_c\,m^3}}.
\qquad
c_m\approx-0.3056,
\label{eq:melnikov}
\end{align}
The leading term follows from the Melnikov method~\cite{melnikov1963stability,guckenheimer2013nonlinear,gao2018self}, and the subleading term is the empirical correction from Ref.~\cite{gao2018self}. The contribution from cluster members is
\begin{align}
\mathcal{L}(R_c)\equiv
\sum_{i\in\text{cluster}}
\frac{d_i}{\sum_jd_j}\,
\frac{
\sqrt{(K_iR_c)^2-(P_i-P_c)^2}
}{
K_iR_c
},
\label{eq:locked}
\end{align}
and the contribution from rotors outside the cluster is
\begin{align}
\mathcal{D}(R_c)\equiv
-\sum_{i\notin\text{cluster}}
\frac{d_i}{\sum_jd_j}\,
\frac{\gamma^2mK_iR_c}
{2[(P_i-P_c)^2m^2+\gamma^4]}.
\label{eq:drift}
\end{align}

Without inertia, the entrainment radius reduces to the phase\-/locking radius,
\begin{align}
r_{c,i}=K_iR_c.
\end{align}
The cluster members then contribute
\begin{align}
\mathcal{L}(R_c)\equiv
\sum_{i\in\text{cluster}}
\frac{d_i}{\sum_jd_j}\,
\frac{
\sqrt{(K_iR_c)^2-(P_i-P_c)^2}
}{
K_iR_c
},
\end{align}
and the drifting contribution vanishes,
\begin{align}
\mathcal{D}(R_c)=0.
\end{align}

For either case, the integrand
\begin{align}
R_c-\mathcal{L}(R_c)-\mathcal{D}(R_c)
\label{eq:integrand}
\end{align}
determines cluster growth. A positive value means that increasing $R_c$ requires external work; a negative value allows spontaneous growth. A crossing from negative to positive identifies a stable cluster.

We define the \Adhoc potential by accumulating the integrand as $R_c$ increases,
\begin{align}
\mathcal{A}(R_c)\equiv
K\!\int_0^{R_c}\!
\big[
R'_c-\mathcal{L}(R'_c)-\mathcal{D}(R'_c)
\big]\,dR'_c.
\label{eq:action}
\end{align}
The potential decreases during spontaneous cluster growth and reaches a local minimum when the integrand vanishes. We denote the nontrivial minimum by $\mathcal{A}(R_c^*)$, with $R_c^*>0$. Its position $R_c^*$ gives the predicted stable cluster order parameter.

To specify the propagation path in Eq.~\eqref{eq:action}, we use the same sampled powers $\{P_i\}$ as in the simulation and discretize
\begin{align}
R_c\in\{0,\delta R,\ldots,1\},
\qquad
\delta R=10^{-4}.
\end{align}
The integration proceeds by increasing $R_c$ one increment at a time.

The construction extends our previous calculation by allowing the cluster center to move. Previously, fixing $P_c=0$ throughout the integration restricted the description to a cluster centered at $P=0$. Here, we update $P_c$ as the cluster grows and follow both $R_c$ and $P_c$ along the propagation path.

At each increment of $R_c$, we update $P_c$ to the mean power of the members from the preceding increment. If the cluster is empty, $P_c$ retains its initial value. We then evaluate the entrainment radii $r_{c,i}$, update cluster membership, and calculate $\mathcal{L}(R_c)$, $\mathcal{D}(R_c)$, and the potential increment.

Explicitly, if $\mathcal{C}_{n-1}$ is the member set at $R_{c,n-1}$,
\begin{align}
P_{c,n}
=
\begin{cases}
P_c(0), & \mathcal{C}_{n-1}=\varnothing,\\[4pt]
\dfrac{1}{|\mathcal{C}_{n-1}|}
\displaystyle\sum_{i\in\mathcal{C}_{n-1}}P_i,
& \mathcal{C}_{n-1}\neq\varnothing.
\end{cases}
\label{eq:Pc_update}
\end{align}
The new member set is
\begin{align}
\mathcal{C}_{n}
=
\left\{
i:
|P_i-P_{c,n}|
<
r_{c,i}(R_{c,n})
\right\}.
\label{eq:cluster_update}
\end{align}
Using $\mathcal{C}_n$, the potential is updated as
\begin{align}
\mathcal{A}(R_{c,n})
=
\mathcal{A}(R_{c,n-1})
+
K
\big[
R_{c,n}
-\mathcal{L}(R_{c,n})
-\mathcal{D}(R_{c,n})
\big]
\delta R.
\label{eq:action_discrete}
\end{align}
Equation~\eqref{eq:action_discrete} is the right\-/endpoint Riemann sum used in the numerical calculation. Repeating the update as $R_c$ increases generates the propagation path.

For the asymmetric distribution, we apply the same procedure after mean\-/centering the powers. The dynamics and entrainment condition use
\begin{align}
P_i=P_i^{\rm raw}-\langle P^{\rm raw}\rangle,
\end{align}
while the relative coupling weights $d_i$ are calculated from the pre\-/shift powers $P_i^{\rm raw}$, as in the simulations.

The two propagation pathways follow the same update procedure and differ only in the initial cluster center $P_c(0)$. For the symmetric distribution, the population\-/driven pathway starts at
\begin{align}
P_c(0)=0,
\end{align}
where $g(P)$ is maximal, whereas the coupling\-/driven pathway starts at
\begin{align}
P_c(0)=P_{\rm hi}.
\end{align}
For the asymmetric distribution, the population\-/driven pathway starts at
\begin{align}
P_c(0)=P_{\rm lo},
\end{align}
whereas the coupling\-/driven pathway starts at
\begin{align}
P_c(0)=P_{\rm hi}.
\end{align}
These initial values specify the seed clusters. Once rotors are entrained, $P_c$ follows the mean power of the cluster members along either pathway.


Updating $P_c$ allows the construction to follow tail\-/seed clusters as they propagate toward the center, extending the fixed\-/center description of central nucleation. The selected pathway agrees with the nucleation strength $h(P)$: population\-/driven for $2\alpha-\lambda<0$ and coupling\-/driven for $2\alpha-\lambda>0$.

\begin{figure*}[!t]
\centering
\includegraphics[width=1.0\linewidth]{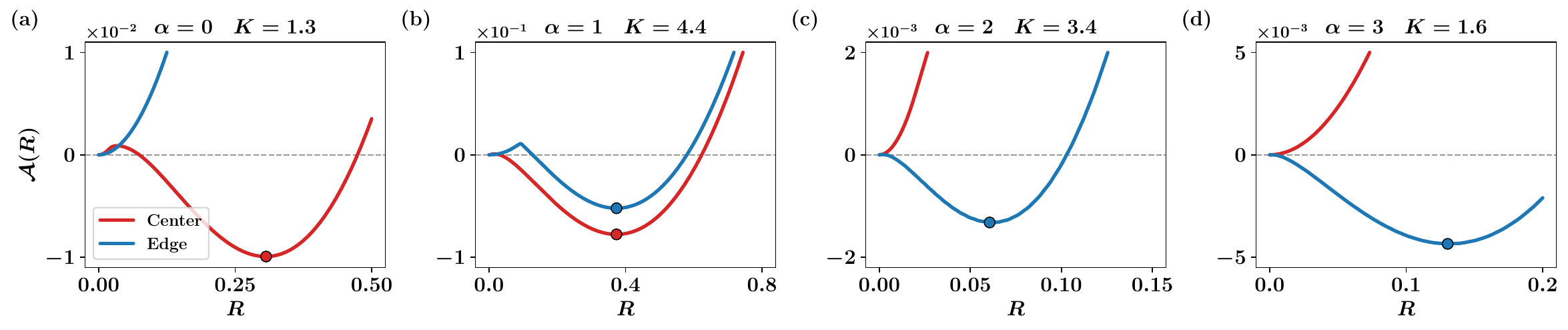}
\caption{\textbf{\Adhoc potential along the two propagation pathways (symmetric distribution, $m=10$).}
(a--d)~$\mathcal{A}(R_c)$ for $\alpha=0,1,2,3$, respectively, at the values of $K$ indicated above the panels. Red and blue curves show the population\-/driven and coupling\-/driven pathways, respectively. Circles mark the potential minima at $R_c^*$. The population\-/driven pathway is deeper for $\alpha=0,1$, and the coupling\-/driven pathway is deeper for $\alpha=2,3$. The gap $\Delta\mathcal{A}$ in (c) is small because $\alpha=2$ lies close to the balance point $\alpha_c=1.5$. The corresponding synchronization results are shown in Fig.~\ref{fig:sym_massive_Rc} of the main text.}
\label{fig:sym_potential}
\end{figure*}

Figure~\ref{fig:sym_potential} shows the potential profiles along the two pathways for the four values of $\alpha$. The population\-/driven pathway is deeper at $\alpha=0,1$, whereas the coupling\-/driven pathway is deeper at $\alpha=2,3$. The resulting order parameters and cluster angular velocities are shown in Fig.~\ref{fig:sym_massive_Rc} of the main text. The asymmetric distribution gives the same ordering across the same balance point [Fig.~\ref{fig:asym_potential}], with a single tail\-/seed cluster instead of a counter\-/rotating pair [Fig.~\ref{fig:asym_massive_Rc}].

\subsection{Scope of the single-cluster description}

Equation~\eqref{eq:action} follows one cluster at a time and neglects inter\-/cluster interactions. We quantify the separation between competing pathways by
\begin{align}
\Delta\mathcal{A}(\alpha)\equiv
\big|
\mathcal{A}(R_c^*)_{\rm selected}
-
\mathcal{A}(R_c^*)_{\rm competing}
\big|.
\label{eq:gap}
\end{align}
A pathway without a nontrivial minimum is assigned $\mathcal{A}(R_c^*)=0$, because that pathway supports no stable cluster.

A large $\Delta\mathcal{A}$ indicates a dominant propagation pathway. The single\-/cluster description is then well controlled because inter\-/cluster interactions have little effect on pathway selection. A small $\Delta\mathcal{A}$ indicates pathways of comparable depth. Clusters associated with both pathways can coexist and interact, making the single\-/cluster description less reliable. Because $\Delta\mathcal{A}\to0$ as $\alpha\to\alpha_c$, the description is least controlled near the balance point.

Near the balance point, the \Adhoc potential still locates the onset and captures the overall shape of $R_c(K)$. However, the predicted and simulated $R_c$ differ more, and nucleation occurs at somewhat larger $K$ than predicted [Fig.~\ref{fig:sym_massive_Rc}(b,c) of the main text]. These deviations occur where population\-/driven and coupling\-/driven clusters coexist and interact. We therefore use $\Delta\mathcal{A}$ to indicate the range in which the single\-/cluster description is quantitatively reliable.

Inter\-/cluster interactions also control the coexistence interval before merger. A large $\Delta\mathcal{A}$ strongly favors one pathway and produces a narrow coexistence interval in $K$; a small $\Delta\mathcal{A}$ makes the pathways comparable and produces a broad interval. The potential gap therefore measures both pathway separation and the validity of the single\-/cluster description.

\clearpage
\newpage

\section{Limit without inertia: symmetric distribution}
\label{sm:noninertial}

\begin{figure*}[h]
\centering
\includegraphics[width=1.0\linewidth]{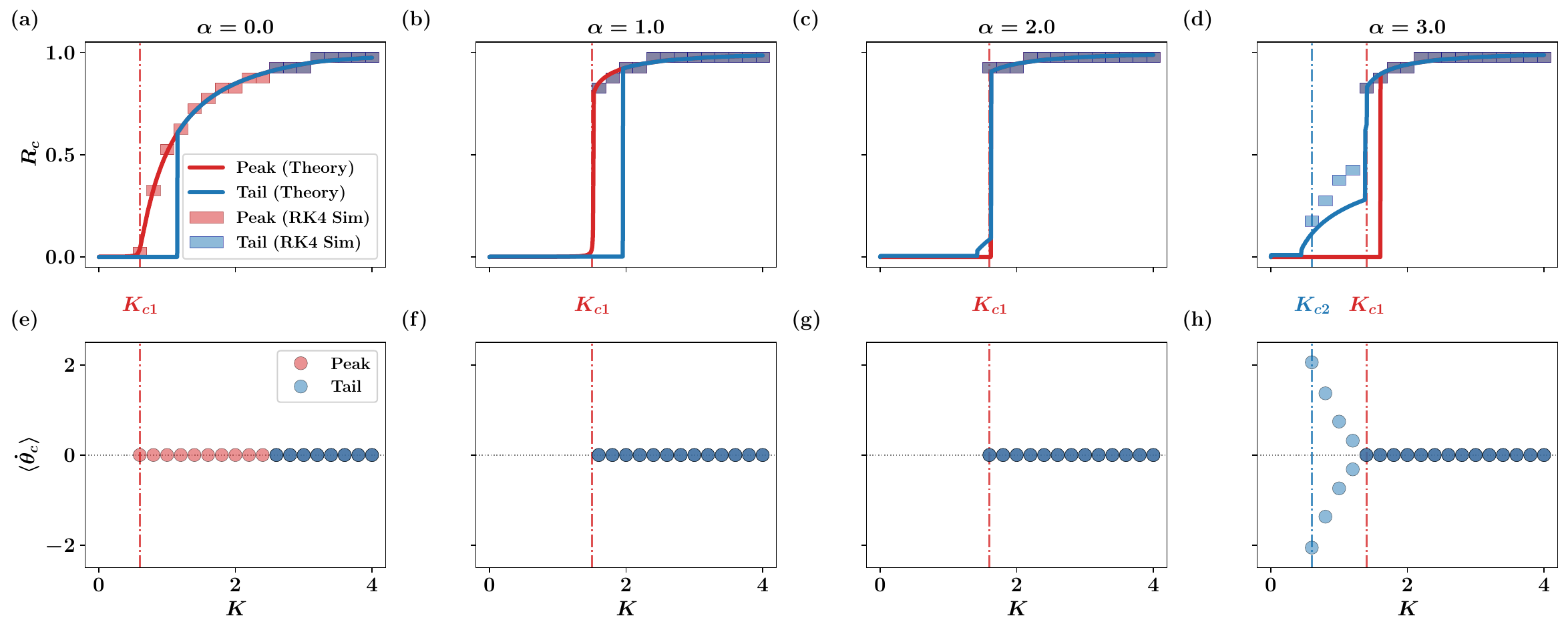}
\caption{\textbf{Seed\-/cluster selection and persistence without inertia ($m=0$).}
Panels correspond to those in Fig.~\ref{fig:sym_massive_Rc} of the main text for $m=10$.
(a--d)~The population\-/driven and coupling\-/driven \Adhoc potential pathways remain distinct, but the merged state occupies only one $R_c$ bin at every $\alpha$, including $\alpha>\alpha_c$, where a counter\-/rotating pair persists for $m=10$.
(e--h)~A cluster at $\langle\dot\theta_c\rangle\approx0$ with $n_c\approx N$ is present above onset. Only for $\alpha=3$ do tail\-/seed clusters retain finite $\pm\langle\dot\theta_c\rangle$ over a finite range of $K$ before merging.}
\label{fig:sym_massless_Rc}
\end{figure*}

Removing inertia preserves the nucleation bins but changes seed\-/cluster persistence. Figure~\ref{fig:sym_massless_Rc} shows the dynamics without inertia for the same $\alpha$, $g(P)$, and $d_i$ as in Fig.~\ref{fig:sym_massive_Rc} of the main text.

The cluster\-/resolved data show the effect of inertia [Fig.~\ref{fig:sym_massless_Rc}]. With inertia, distinct seed clusters coexist over a range of $K$. Without inertia, tail\-/seed clusters persist only briefly before merging into a cluster at $\langle\dot\theta_c\rangle\approx0$ with $n_c\approx N$, leaving one occupied $R_c$ bin. First\-/order Kuramoto models with correlated frequencies and coupling strengths can support traveling\-/wave or standing\-/wave states under specific conditions~\cite{iatsenko2013stationary,xu2016synchronization,bi2017nontrivial}.

Without inertia, rotors rapidly follow the mean field and distinct seed clusters merge soon after nucleation. With inertia, rotors retain their angular velocities and seed clusters persist~\cite{kim2025cluster,kim2026paths}.

\clearpage
\newpage

\section{Asymmetric distribution: generality of the mechanism}
\label{sm:asym}

\begin{figure*}[h]
\centering
\includegraphics[width=1.0\linewidth]{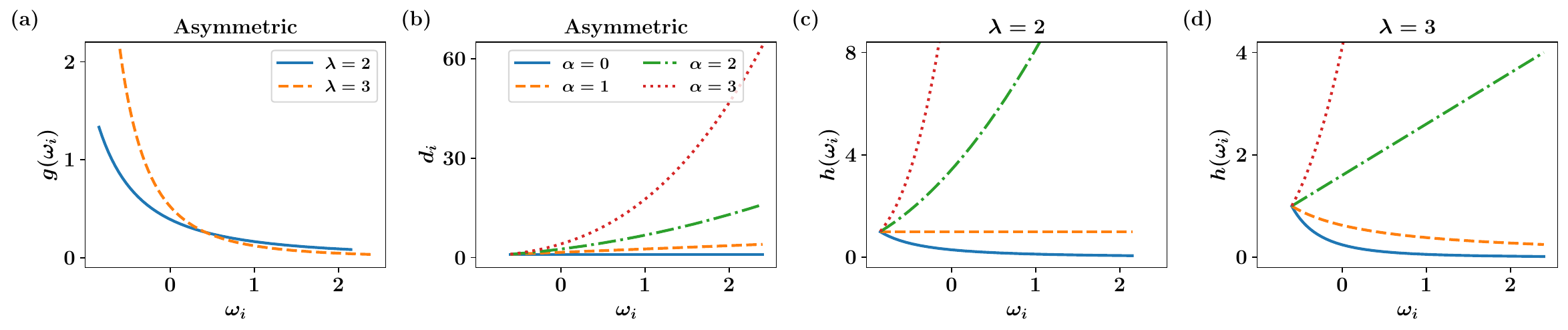}
\caption{\textbf{Rotor population, coupling\-/weight profile, and nucleation strength versus power $P$ (asymmetric distribution).}
(a)~One\-/sided power distribution $g(P)$ for $\lambda=2,3$, peaked at the lower edge.
(b)~Coupling\-/weight profile $d(P)$ for $\alpha=0,1,2,3$, increasing toward the upper tail.
(c,d)~Nucleation strength $h(P)$ for $\lambda=2$ (c) and $\lambda=3$ (d). For $\alpha<\lambda/2$, $h(P)$ peaks at the lower edge (population\-/driven); for $\alpha>\lambda/2$, $h(P)$ increases toward the upper tail (coupling\-/driven). At $\alpha_c=\lambda/2$, the two contributions balance.}
\label{fig:asym_dist}
\end{figure*}

We repeat the analysis for an asymmetric power distribution. Raw powers $P_i^{\rm raw}$ are sampled from $(P^{\rm raw}+P_0)^{-\lambda}$ on $[0,P_b]$ and mean\-/centered to obtain $P_i$, while $d_i$ is computed from the pre\-/shift powers. The resulting $g(P)$ peaks at the lower edge, and $d(P)$ increases toward the upper tail [Fig.~\ref{fig:asym_dist}]. The balance point remains $\alpha_c=\lambda/2$. A single seed cluster nucleates near the lower edge in the population\-/driven regime and near the upper tail in the coupling\-/driven regime. Both pathways then grow by accretion because only one seed cluster nucleates. We use $m=10$ unless stated otherwise.

\clearpage
\newpage

\subsection{Nucleation bin and persistence}

\begin{figure*}[h]
\centering
\includegraphics[width=1.0\linewidth]{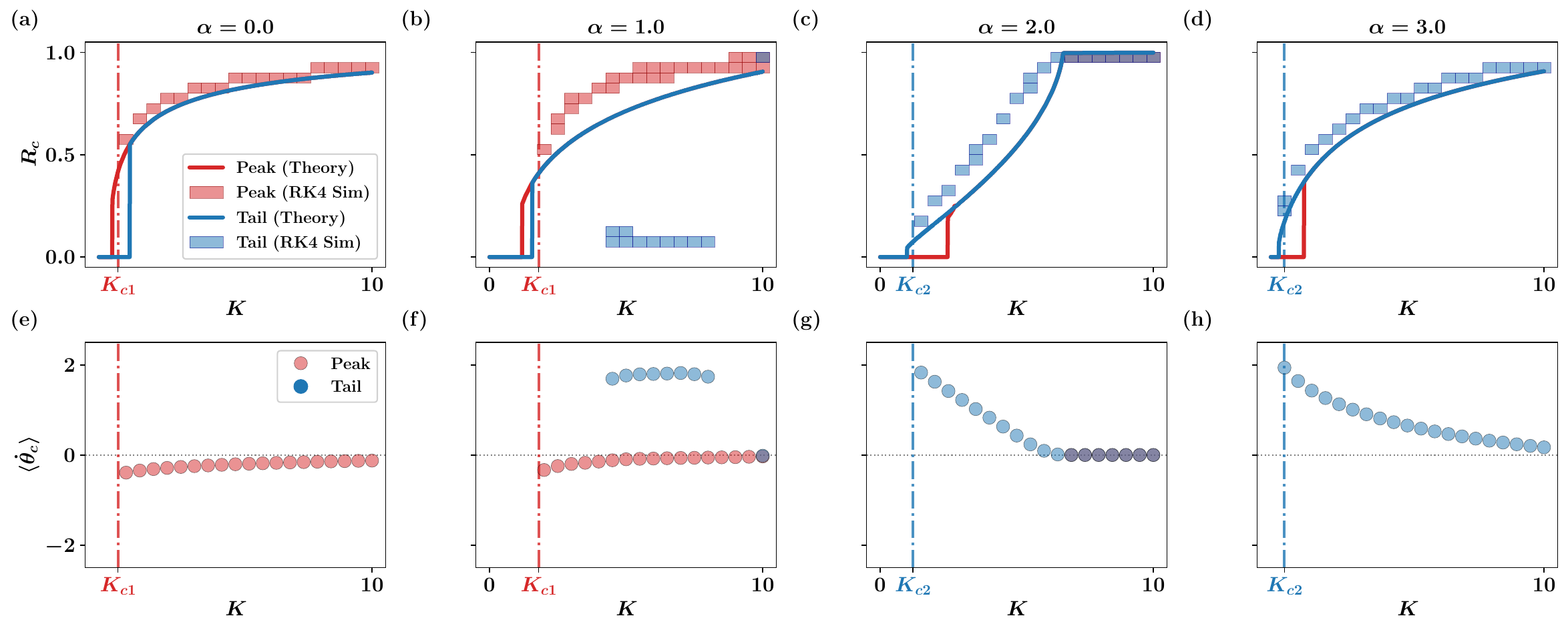}
\caption{\textbf{Population\-/driven and coupling\-/driven synchronization (asymmetric distribution, $m=10$).}
Panels correspond to those in Fig.~\ref{fig:sym_massive_Rc} of the main text.
(a--d)~$R_c$ versus $K$. Solid curves: population\-/driven (red) and coupling\-/driven (blue) pathways predicted by the \Adhoc potential accumulated along each path. Squares: the most populated $R_c$ bins over $64$ realizations at each $K$. Dash\-/dotted vertical lines: predicted onsets $K_{c1}$ (red) and $K_{c2}$ (blue).
(e--h)~Modal positions of the mean cluster angular velocities $\langle\dot\theta_c\rangle$. For $\alpha<\alpha_c$, a population\-/driven seed cluster nucleates near the lower edge; for $\alpha>\alpha_c$, a coupling\-/driven seed cluster nucleates near the upper tail. In both cases, $\langle\dot\theta_c\rangle$ approaches zero as the cluster grows.
The \Adhoc potential reproduces $R_c(K)$ and selects the pathway observed in the simulations. Agreement is strongest for $\alpha=0,3$ and weakest for $\alpha=1,2$, where nucleation occurs at somewhat larger $K$ than predicted. The corresponding potential profiles are shown in Fig.~\ref{fig:asym_potential}.}
\label{fig:asym_massive_Rc}
\end{figure*}

The cluster\-/resolved analysis identifies the nucleation bin [Fig.~\ref{fig:asym_massive_Rc}]. In the population\-/driven regime ($\alpha=0,1$), a seed cluster nucleates near the lower edge at $K_{c1}$ and grows by accretion. In the coupling\-/driven regime ($\alpha=2,3$), a seed cluster nucleates near the upper tail at $K_{c2}$ and grows by accretion. In both regimes, the cluster grows continuously toward $n_c\approx N$, and its angular velocity approaches zero as its mean power approaches the mean\-/centered value.

\begin{figure*}[h]
\centering
\includegraphics[width=1.0\linewidth]{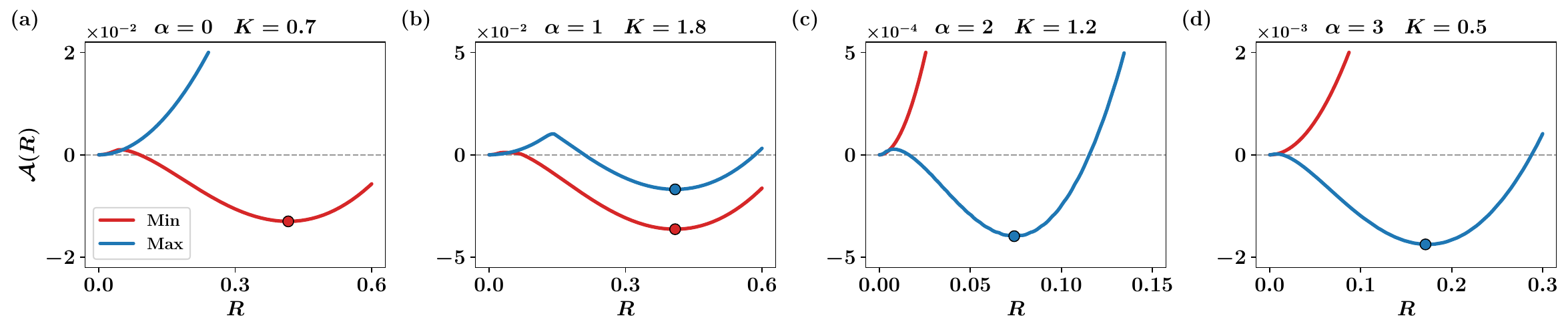}
\caption{\textbf{\Adhoc potential along the two propagation pathways (asymmetric distribution, $m=10$).}
(a--d)~$\mathcal{A}(R_c)$ for $\alpha=0,1,2,3$, respectively, at the values of $K$ where the selected cluster stabilizes, indicated above the panels. Red and blue curves show the population\-/driven and coupling\-/driven pathways, respectively. Circles mark the minima at $R_c^*$. The deeper pathway is selected: population\-/driven for $\alpha=0,1$ and coupling\-/driven for $\alpha=2,3$. The gap $\Delta\mathcal{A}$ is large for $\alpha=0,3$ and small for $\alpha=1,2$, indicating stronger competition near the balance point. The corresponding synchronization results are shown in Fig.~\ref{fig:asym_massive_Rc}.}
\label{fig:asym_potential}
\end{figure*}

The same \Adhoc potential method favors the population\-/driven pathway for $\alpha=0,1$ and the coupling\-/driven pathway for $\alpha=2,3$ [Fig.~\ref{fig:asym_potential}]. The prediction of coupling\-/driven nucleation away from the maximum of $g(P)$ confirms that the construction applies to asymmetric distributions. The potential gap $\Delta\mathcal{A}$ is large for $\alpha=0,3$ and small for $\alpha=1,2$, consistent with stronger competition near the balance point.

\clearpage
\newpage

\subsection{Transient time}

\begin{figure*}[h]
\centering
\includegraphics[width=1.0\linewidth]{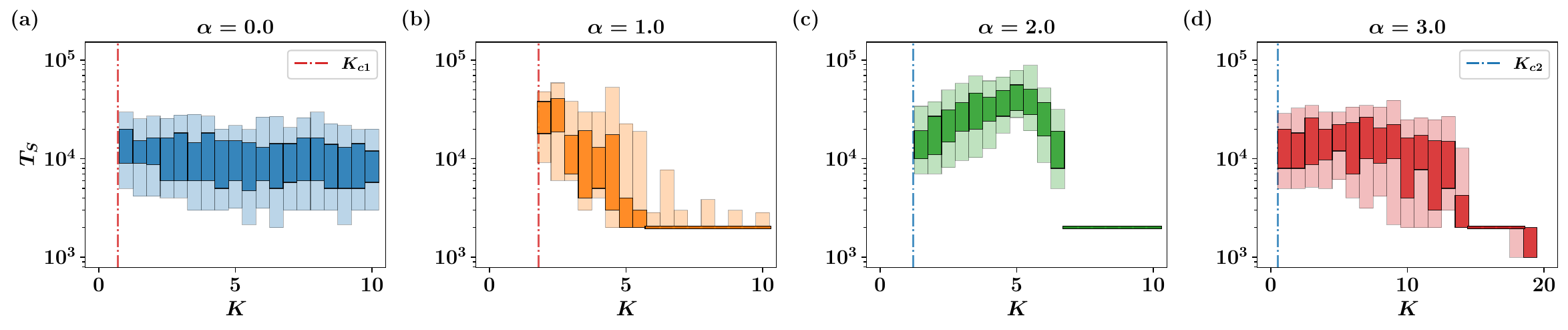}
\caption{\textbf{Transient\-/time distributions for accretion (asymmetric distribution, $m=10$).}
Panels and box construction are as in Fig.~\ref{fig:critical}.
(a--d)~$T_S$ over $64$ realizations versus $K$. Dash\-/dotted vertical lines: measured onsets $K_{c1}$ (red) and $K_{c2}$ (blue). Population\-/driven regime ($\alpha=0,1$): $T_S$ is large and broadly distributed just above $K_{c1}$ and decreases as $K$ increases. Coupling\-/driven regime ($\alpha=2,3$): $T_S$ is large and broadly distributed just above $K_{c2}$ and decreases as $K$ increases. Both regimes grow by accretion because only one seed cluster nucleates; the merger observed for the symmetric coupling\-/driven case is absent.}
\label{fig:asym_critical}
\end{figure*}

Propagation proceeds by accretion in both regimes [Fig.~\ref{fig:asym_critical}]. A single seed cluster nucleates at $K_{c1}$ in the population\-/driven regime or at $K_{c2}$ in the coupling\-/driven regime, then entrains the remaining rotors one at a time. Immediately above either onset, $T_S$ is large and broadly distributed. As $K$ increases, entrainment accelerates and both $T_S$ and its spread decrease. The pathways therefore differ in nucleation bin and onset coupling, while sharing the same growth mechanism.

\clearpage
\newpage

\subsection{Robustness and resilience of the synchronized state}

\begin{figure*}[h]
\centering
\includegraphics[width=1.0\linewidth]{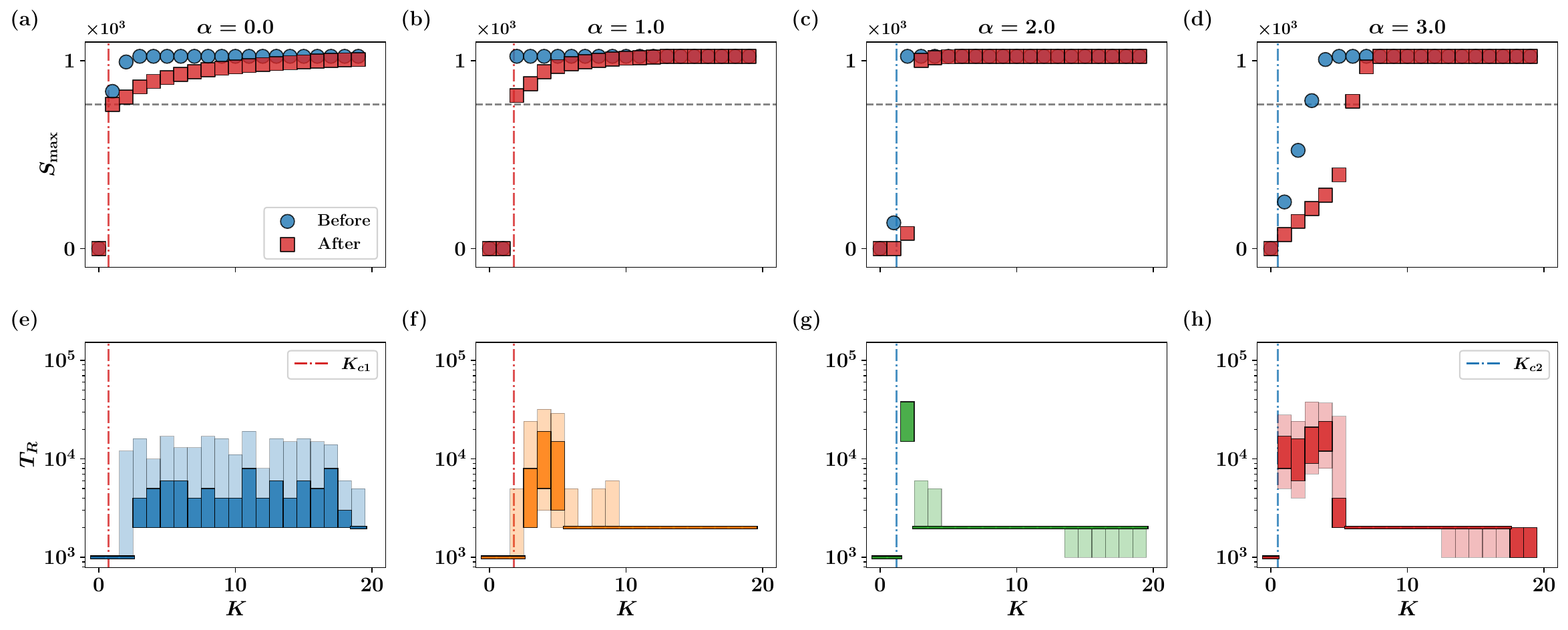}
\caption{\textbf{Robustness and resilience of the synchronized state (asymmetric distribution, $m=10$).}
(a--d)~Largest cluster size $S_{\max}$ before (blue) and after (red) the perturbation; medians over $64$ realizations. Population\-/driven regime ($\alpha=0,1$): a macroscopic core survives and recovery proceeds by accretion. Coupling\-/driven regime ($\alpha=2,3$): the detached tail rotors nucleate a new cluster and recovery proceeds by merger.
(e--h)~Recovery time $T_R$, with measured onsets $K_{c1}$ (red) and $K_{c2}$ (blue). Population\-/driven recovery by accretion gives relatively long, broadly distributed recovery times. Successful coupling\-/driven recovery by merger at strong coupling gives shorter, more narrowly distributed recovery times.}
\label{fig:asym_stability}
\end{figure*}

Recovery follows the same two mechanisms as for the symmetric distribution [Fig.~\ref{fig:asym_stability}]. The perturbation detaches the high-$d_i$ tail rotors. In the population\-/driven regime, these rotors lie at the margin of the macroscopic cluster, so the core survives and recovery proceeds by accretion. In the coupling\-/driven regime, these rotors are the original tail\-/seed rotors. The detached rotors nucleate a new cluster comparable to the surviving core, so recovery proceeds by merger.

The recovery mechanisms produce distinct $S_{\max}$ and $T_R$. Population\-/driven recovery preserves a macroscopic core at all $K$ above onset but restores detached rotors one at a time, giving relatively long recovery times. Coupling\-/driven recovery fails to preserve a macroscopic cluster at weak coupling but rapidly restores the unperturbed $S_{\max}$ once coupling is strong enough for merger. A broad distribution of $T_R$ marks the range where the two clusters coexist without merging.

The asymmetric distribution reproduces the selection criterion and recovery mechanisms across the same balance point. Its propagation pathways differ in nucleation bin and onset coupling, but both grow by accretion because only one seed cluster nucleates.

\clearpage
\newpage

\subsection{Limit without inertia: asymmetric distribution}

\begin{figure*}[h]
\centering
\includegraphics[width=1.0\linewidth]{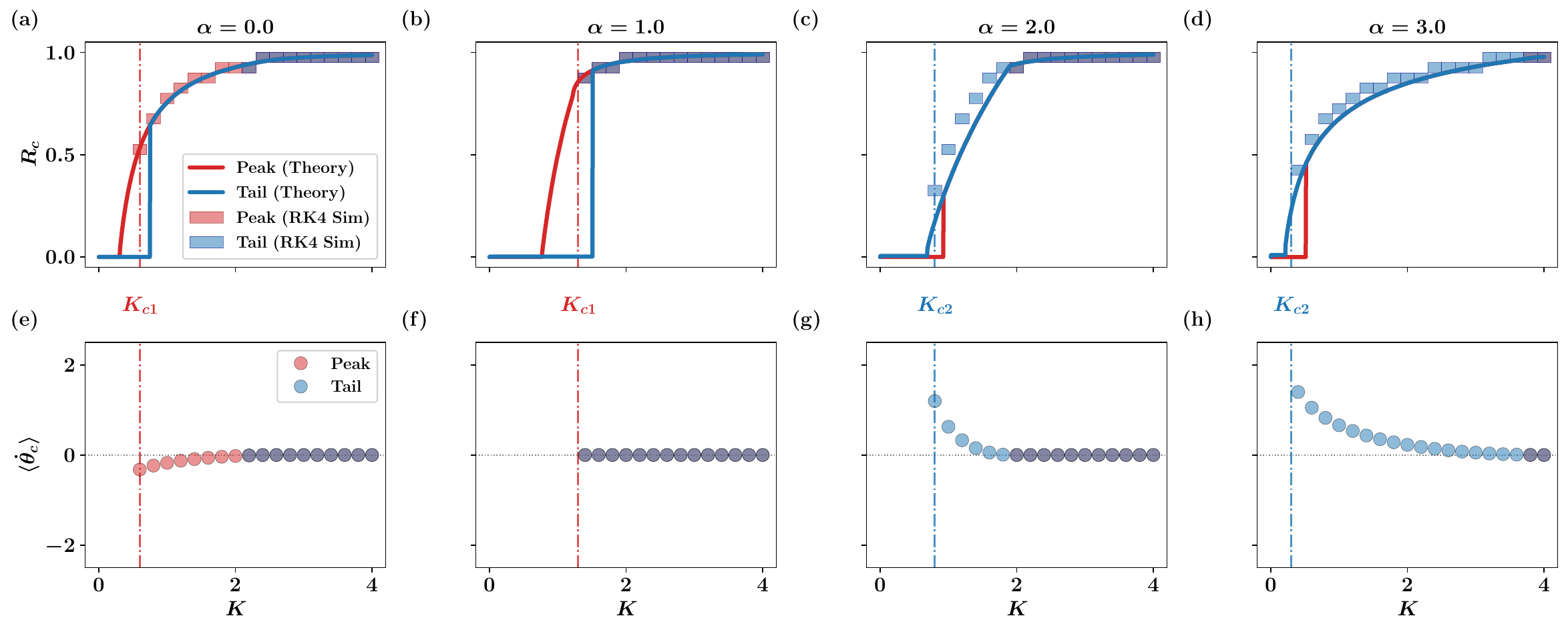}
\caption{\textbf{Same selection without inertia (asymmetric distribution, $m=0$).}
Panels correspond to those in Fig.~\ref{fig:asym_massive_Rc} for $m=10$.
(a--d)~The seed cluster nucleates in the bin identified by $h(P)$ and grows toward $n_c\approx N$, as for $m=10$.
(e--h)~For $\alpha<\alpha_c$, the population\-/driven seed cluster nucleates near the lower edge; for $\alpha>\alpha_c$, the coupling\-/driven seed cluster nucleates near the upper tail. In both regimes, $\langle\dot\theta_c\rangle$ approaches zero as the cluster grows.}
\label{fig:asym_massless_Rc}
\end{figure*}

The same selection rule holds without inertia [Fig.~\ref{fig:asym_massless_Rc}]. A seed cluster nucleates near the lower edge for $\alpha<\alpha_c$ and near the upper tail for $\alpha>\alpha_c$, then grows toward $n_c\approx N$. The nucleation bin is therefore determined by $g(P)$ and $d(P)$, while inertia controls persistence. Removing inertia does not qualitatively change the propagation mechanism because only one seed cluster nucleates for the asymmetric distribution.

\clearpage
\onecolumngrid

\section{Data-derived annealed representations of European grids}
\label{sm:grids}

We test the selection\=/persistence mechanism for more complex power distributions and coupling\-/weight profiles using the French, German, Spanish, and British subgraphs of the PanTaGruEl model of the European transmission system~\cite{pagnier2019inertia}. Each national subgraph is mapped to an all\-/to\-/all representation. ``European grids'' subsection describes the mapping, rescaling, and restoration of power balance after removing cross\-/border lines. We set $\gamma=1$ for every bus and compare $m_i=0$ (without inertia) with $m_i=10$ (with inertia).

The four grids show qualitatively similar power distributions and power\=/coupling correlations [Figs.~\ref{fig:fr_data_sm}, \ref{fig:de_data}, \ref{fig:es_data}, and~\ref{fig:uk_data}]. In the corresponding annealed representations [Figs.~\ref{fig:grid}, \ref{fig:grid_DE}, \ref{fig:grid_ES}, and~\ref{fig:grid_UK}], consumers form a population peak near $P=0$, while generators extend into a high\-/power tail. Together, $g(P)$ and $d(P)$ determine the nucleation strength $h(P)$, which has maxima near the consumer peak and in the generator tail. These maxima identify the nucleation bins for population\-/driven and coupling\-/driven seed clusters, respectively.

Inertia controls the persistence of the resulting seed clusters. Without inertia, the generator\-/tail cluster has a finite $\langle\dot\theta_c\rangle$ at weak coupling and merges with the consumer cluster at intermediate $K$, forming a single cluster with $\langle\dot\theta_c\rangle\approx0$. With inertia, the generator\-/tail cluster retains a finite $\langle\dot\theta_c\rangle$ and coexists with the consumer cluster over a broad range of $K$. All four annealed representations therefore exhibit the same selection\=/persistence mechanism.

For each country, the first figure shows the power distribution and total coupling strength, and the second shows synchronization without and with inertia. All grid\-/data figures follow the plotting conventions of Fig.~\ref{fig:fr_data} of the main text; the French data are repeated here for comparison. In the annealed figures, $P_i$ is the rescaled power and $K$ is the global coupling strength. Panels (e,f) show the modal positions of the mean cluster angular velocities. Red and blue curves in panels (c--f) give the \Adhoc potential estimates for the population\-/driven and coupling\-/driven pathways, respectively.

\clearpage
\twocolumngrid

\begingroup
\raggedbottom
\setlength{\intextsep}{6pt}
\setlength{\abovecaptionskip}{3pt}
\setlength{\belowcaptionskip}{0pt}

\subsection{French grid}
\label{sm:french_grid}

\begin{figure}[H]
\centering
\includegraphics[width=1.0\linewidth]{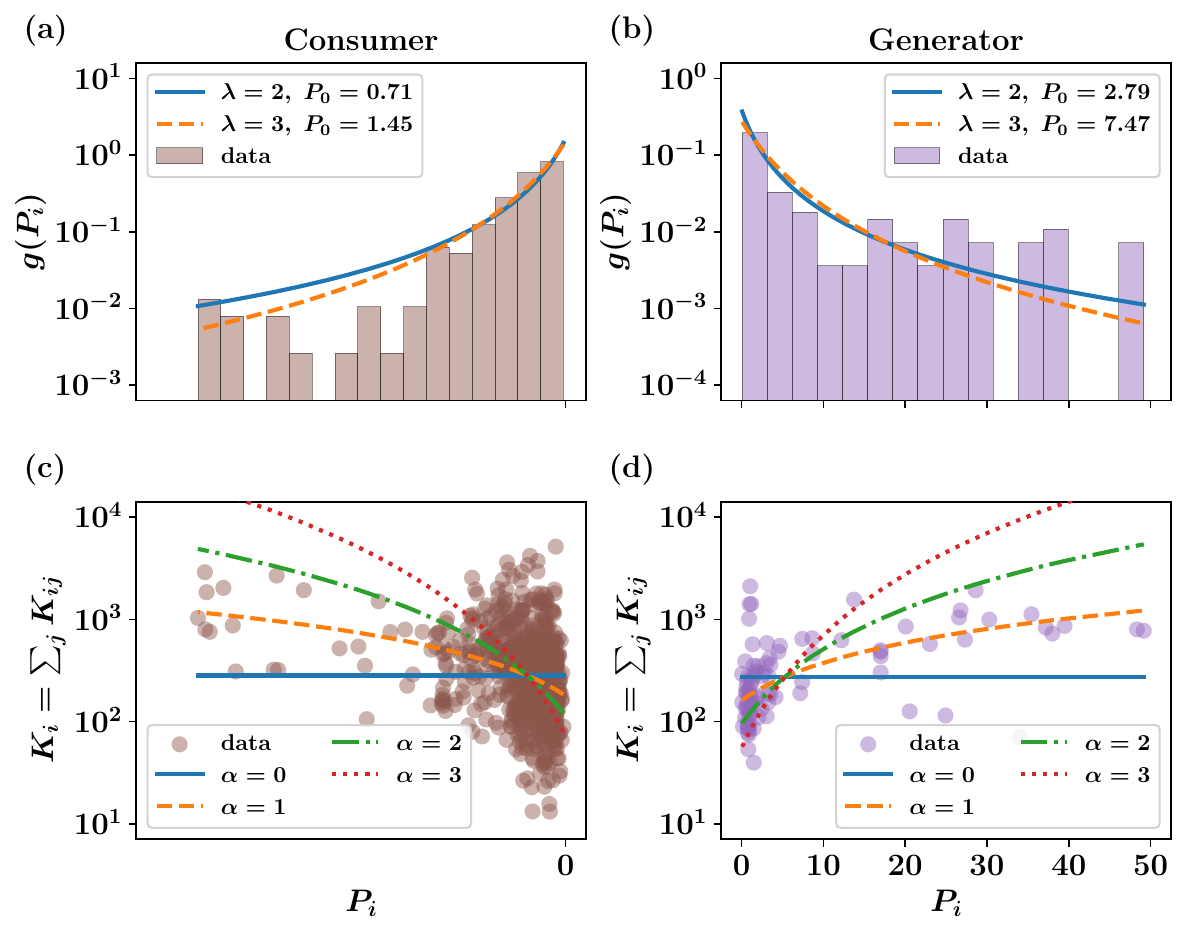}
\caption{\textbf{French grid data.} (a,b)~$g(P)$; (c,d)~$K_i$ versus $P_i$. Consumers: (a,c); generators: (b,d).}
\label{fig:fr_data_sm}
\end{figure}

\begin{figure}[H]
\centering
\includegraphics[width=1.0\linewidth]{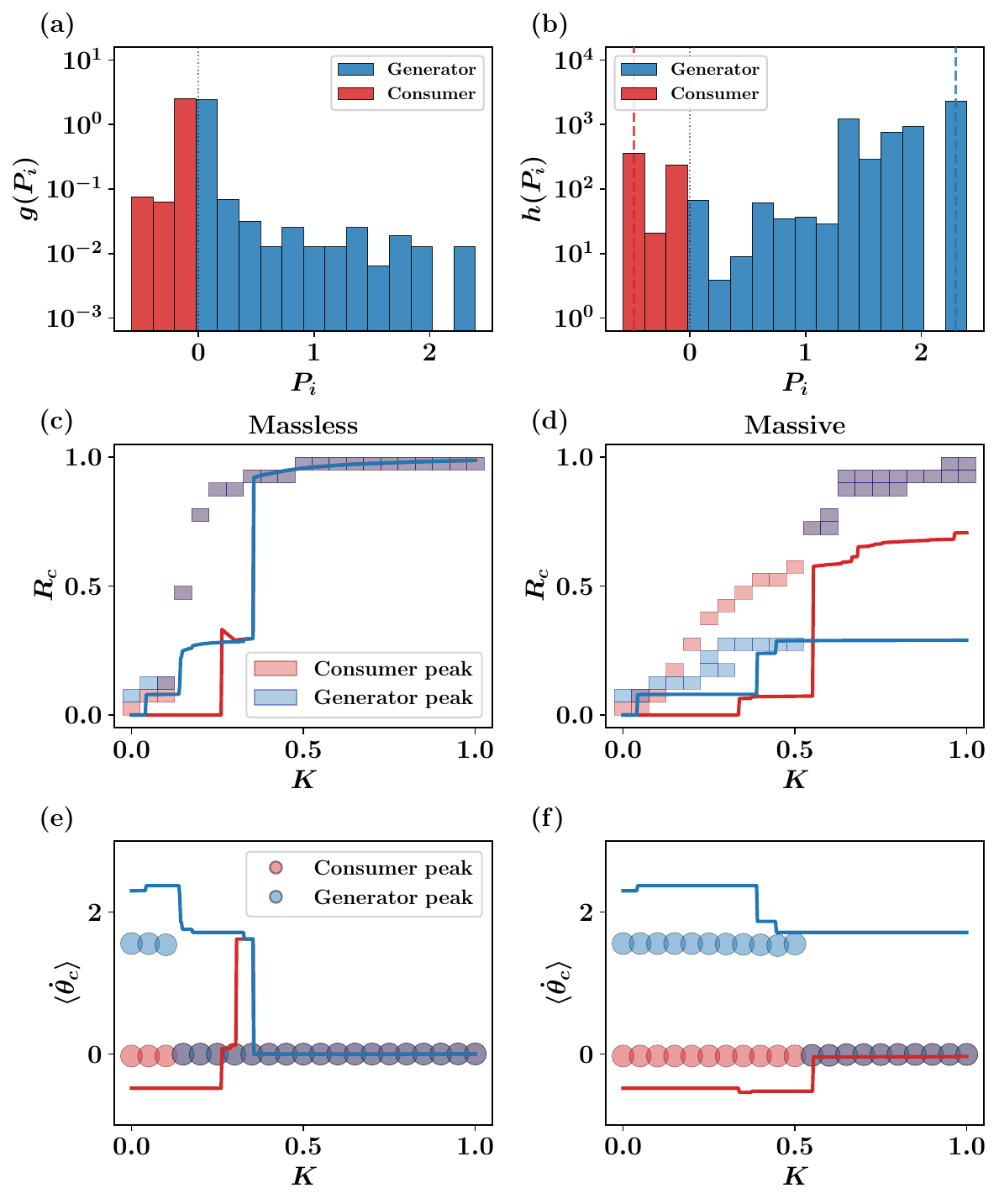}
\caption{\textbf{Annealed French grid.} (a)~$g(P)$; (b)~$h(P)$; (c,d)~$R_c$; (e,f)~$\langle\dot\theta_c\rangle$. Left: $m=0$; right: $m=10$.}
\label{fig:grid}
\end{figure}

\newpage

\subsection{German grid}
\label{sm:german_grid}

\begin{figure}[H]
\centering
\includegraphics[width=1.0\linewidth]{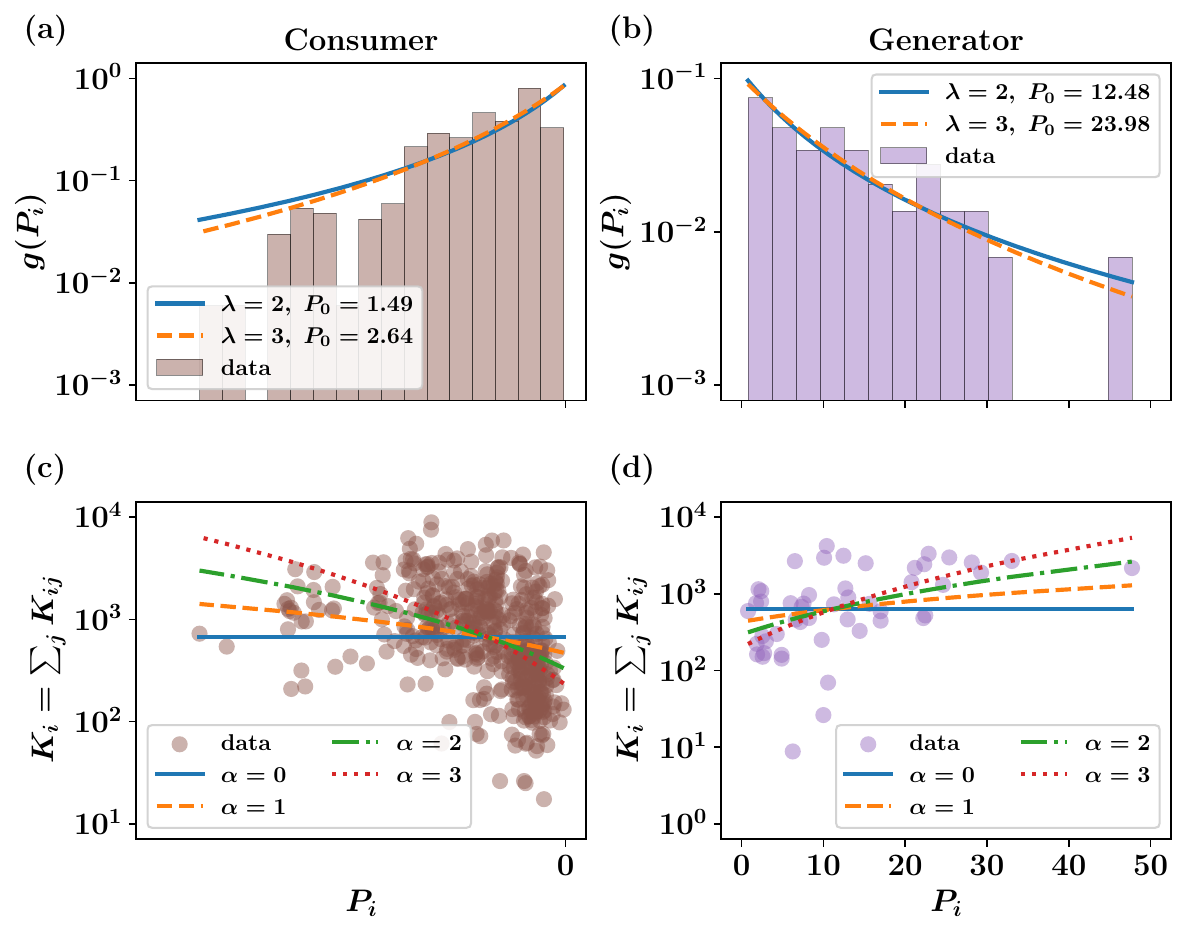}
\caption{\textbf{German grid data.} (a,b)~$g(P)$; (c,d)~$K_i$ versus $P_i$. Consumers: (a,c); generators: (b,d).}
\label{fig:de_data}
\end{figure}

\begin{figure}[H]
\centering
\includegraphics[width=1.0\linewidth]{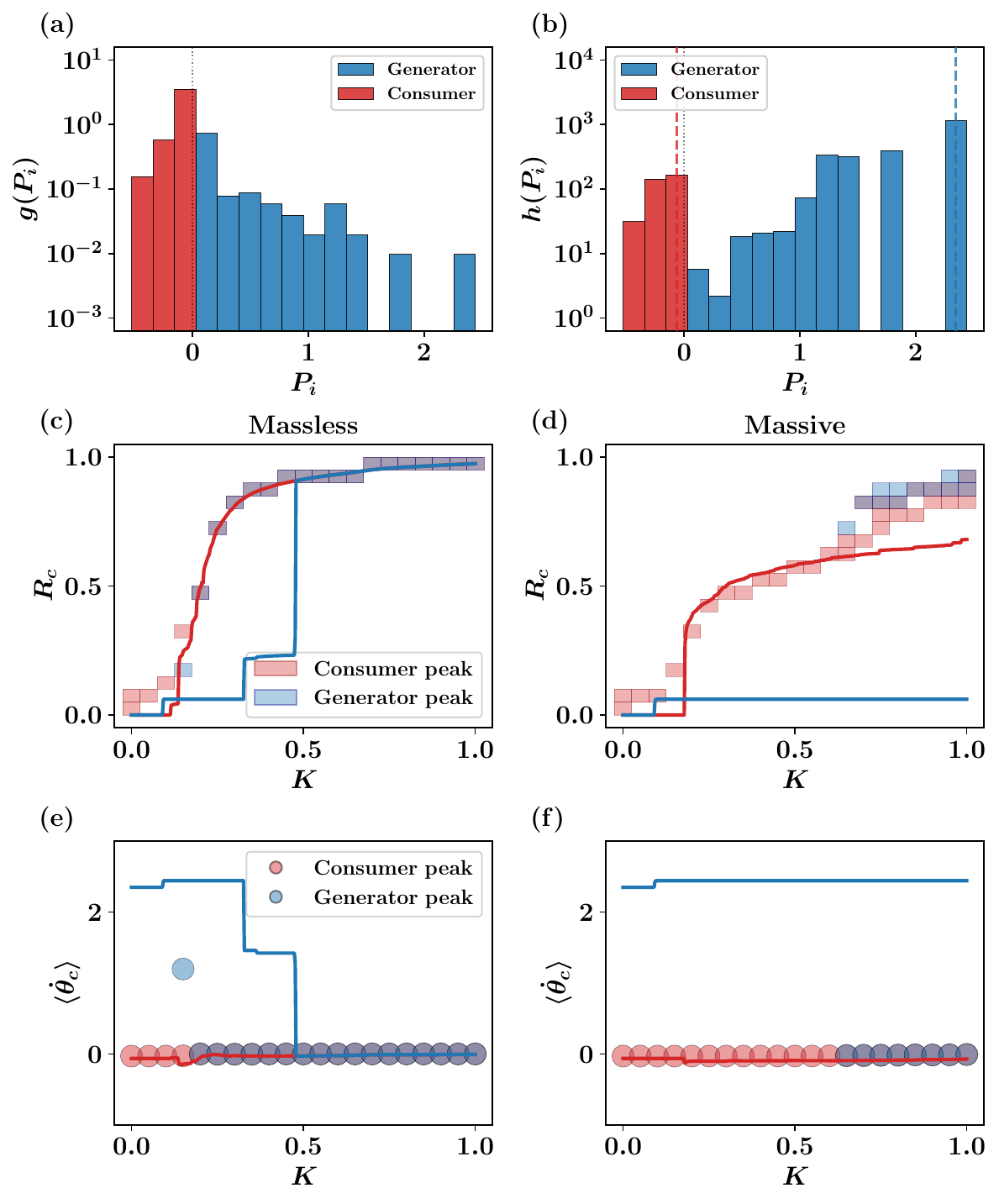}
\caption{\textbf{Annealed German grid.} (a)~$g(P)$; (b)~$h(P)$; (c,d)~$R_c$; (e,f)~$\langle\dot\theta_c\rangle$. Left: $m=0$; right: $m=10$.}
\label{fig:grid_DE}
\end{figure}

\clearpage

\subsection{Spanish grid}
\label{sm:spanish_grid}

\begin{figure}[H]
\centering
\includegraphics[width=1.0\linewidth]{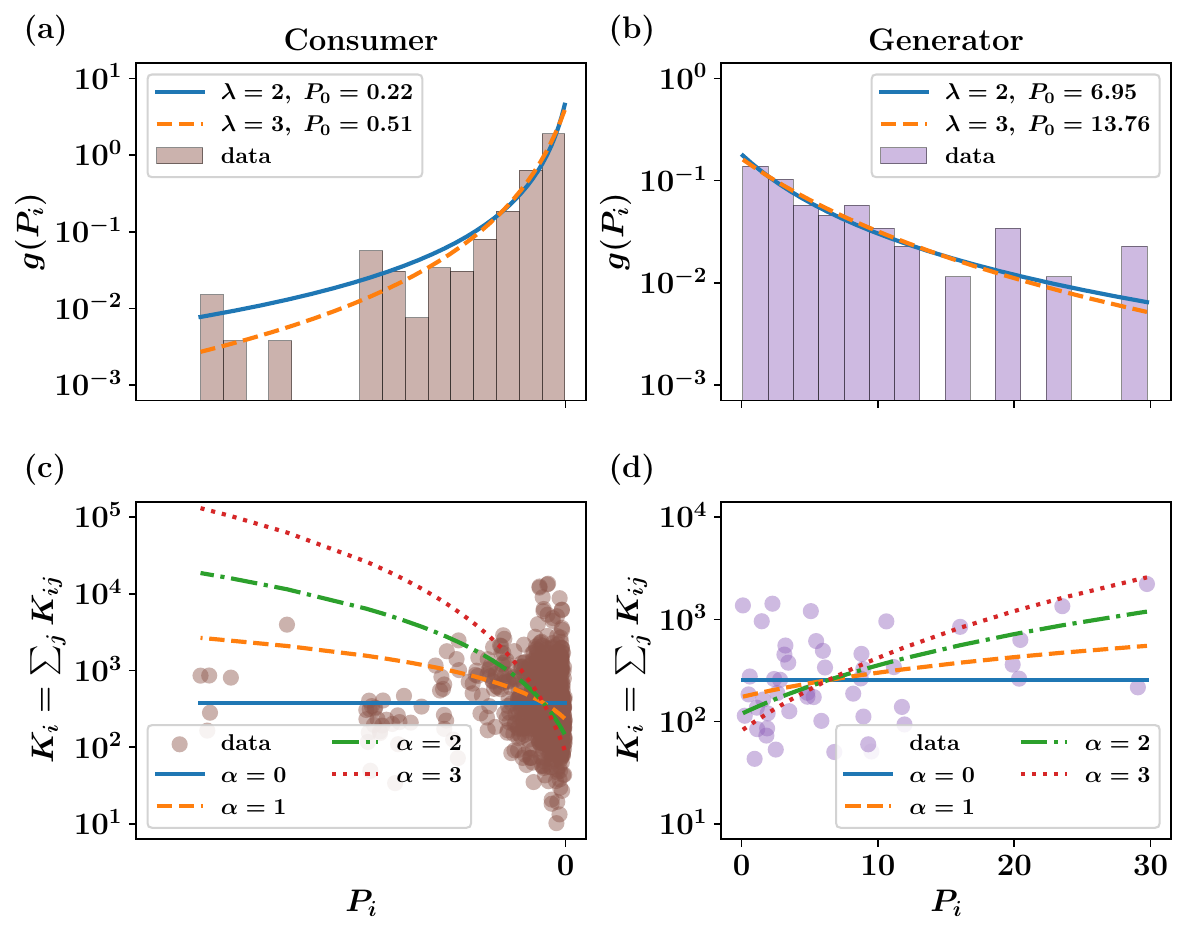}
\caption{\textbf{Spanish grid data.} (a,b)~$g(P)$; (c,d)~$K_i$ versus $P_i$. Consumers: (a,c); generators: (b,d).}
\label{fig:es_data}
\end{figure}

\begin{figure}[H]
\centering
\includegraphics[width=1.0\linewidth]{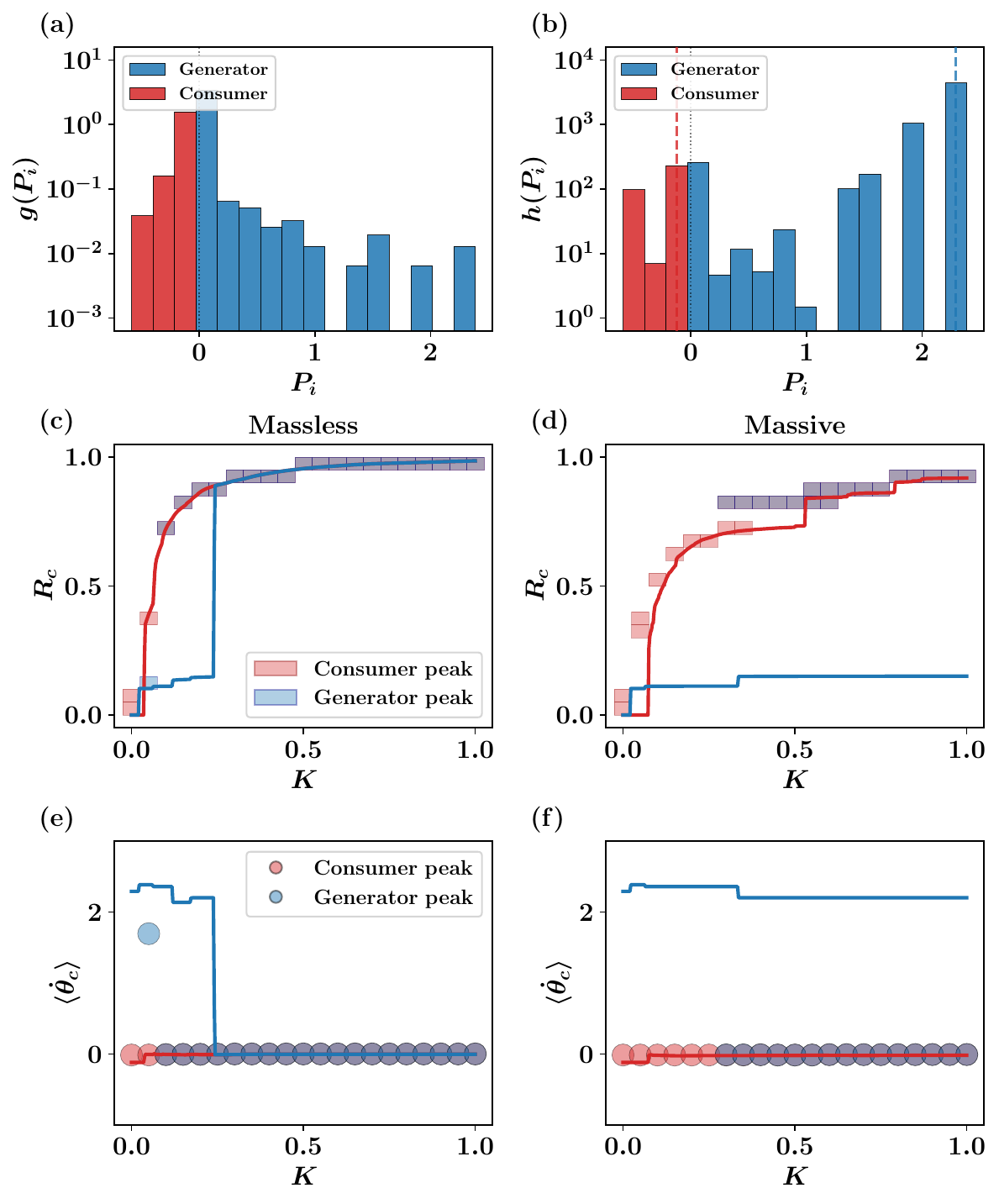}
\caption{\textbf{Annealed Spanish grid.} (a)~$g(P)$; (b)~$h(P)$; (c,d)~$R_c$; (e,f)~$\langle\dot\theta_c\rangle$. Left: $m=0$; right: $m=10$.}
\label{fig:grid_ES}
\end{figure}

\newpage

\subsection{British grid}
\label{sm:british_grid}

\begin{figure}[H]
\centering
\includegraphics[width=1.0\linewidth]{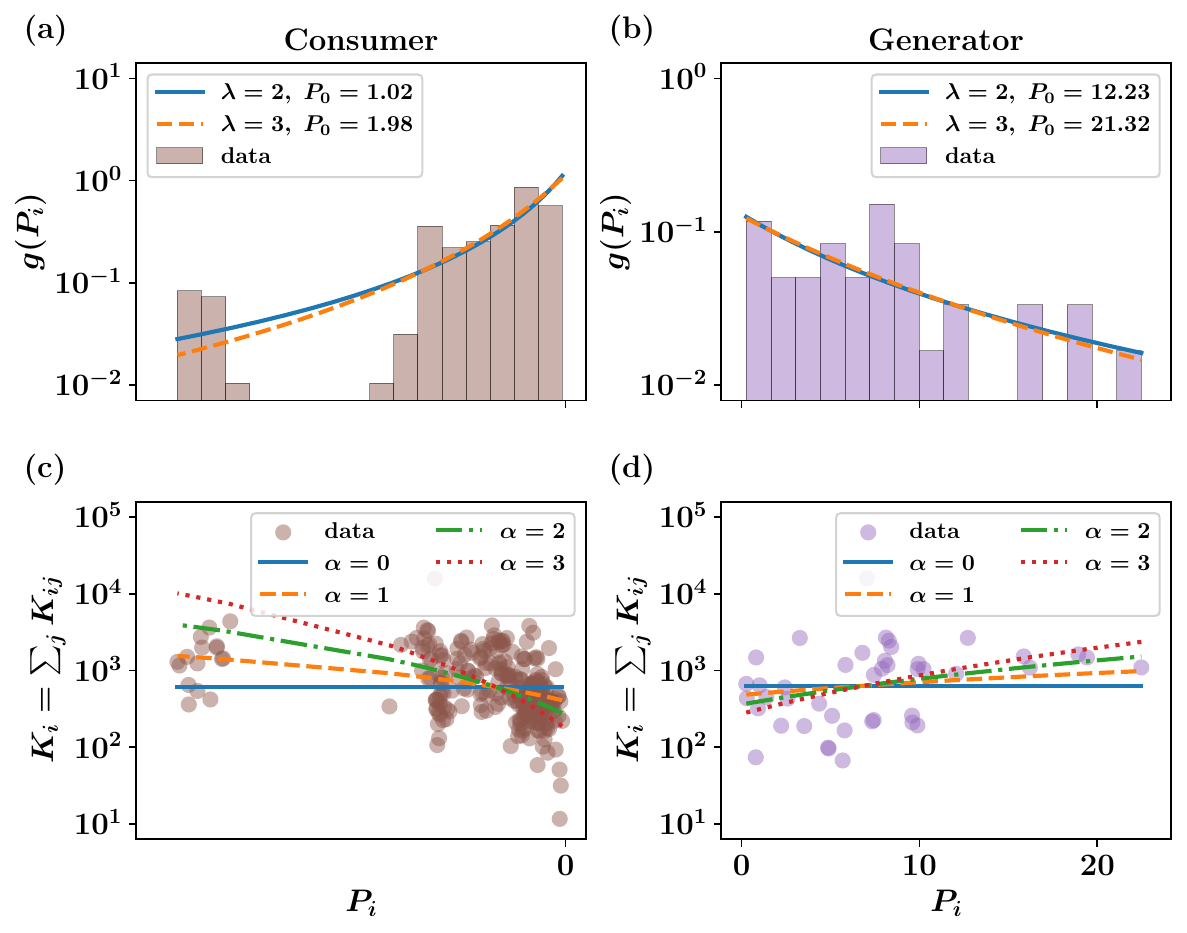}
\caption{\textbf{British grid data.} (a,b)~$g(P)$; (c,d)~$K_i$ versus $P_i$. Consumers: (a,c); generators: (b,d).}
\label{fig:uk_data}
\end{figure}

\begin{figure}[H]
\centering
\includegraphics[width=1.0\linewidth]{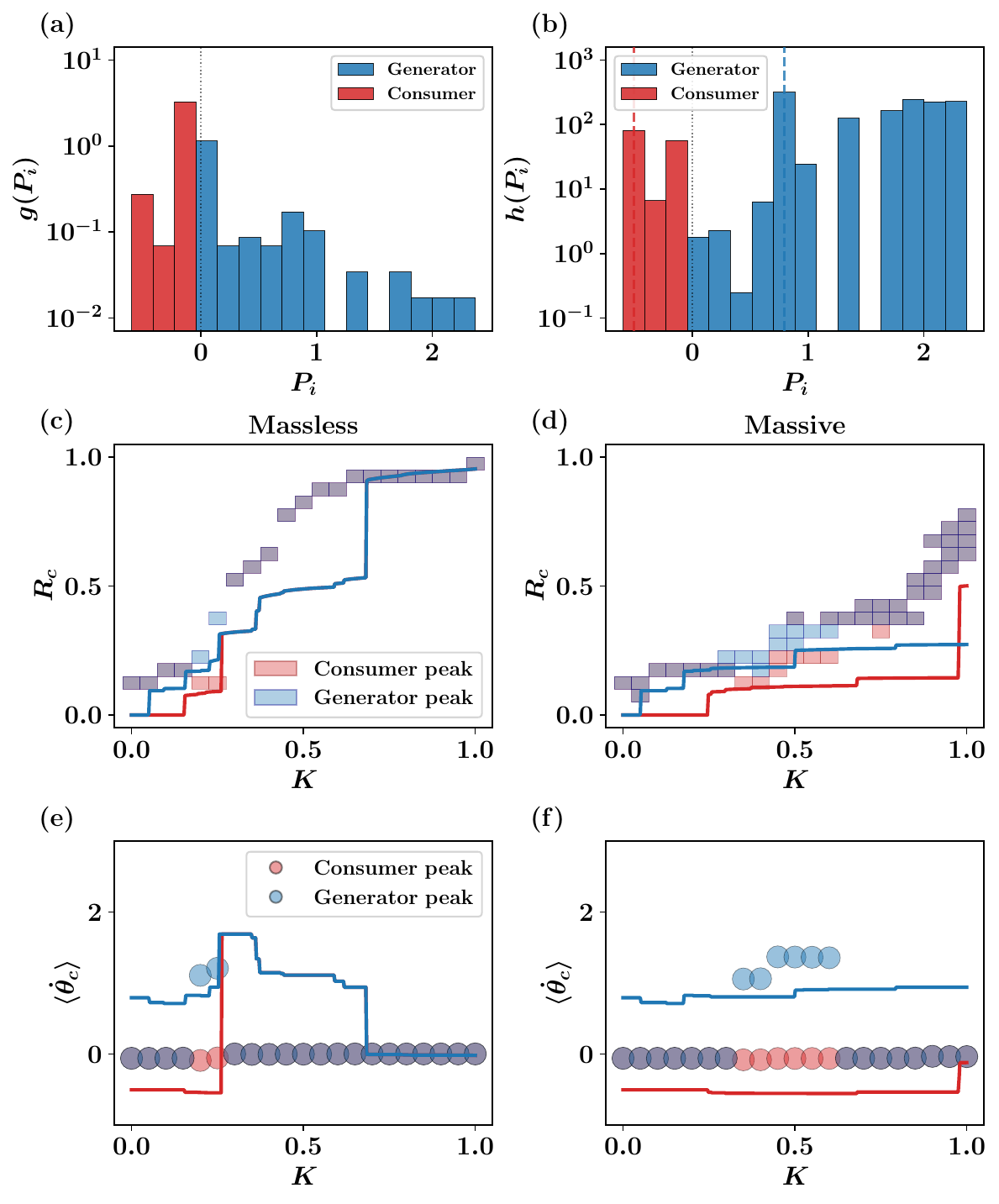}
\caption{\textbf{Annealed British grid.} (a)~$g(P)$; (b)~$h(P)$; (c,d)~$R_c$; (e,f)~$\langle\dot\theta_c\rangle$. Left: $m=0$; right: $m=10$.}
\label{fig:grid_UK}
\end{figure}

\clearpage
\endgroup
\onecolumngrid

\renewcommand{\refname}{Supplementary References}
\putbib
\end{bibunit}

\end{document}